\documentclass{jfm}
\usepackage{graphicx}
\usepackage{epstopdf, epsfig}
\usepackage{natbib}
\usepackage{amsmath}
\usepackage{amssymb}
\usepackage{soul}
\usepackage[percent]{overpic}
\usepackage[colorlinks=true,linkcolor=blue]{hyperref}%

\newcommand{\gradp}{\boldsymbol{\nabla} p}

\usepackage{color}
\definecolor{mycolor}{rgb}{0.2,0.2,1.0}
\shorttitle{Scaling laws for jets of single cavitation bubbles}
\shortauthor{O. Supponen et al.}

\title{Scaling laws for jets of single cavitation bubbles} 

\author{Outi Supponen\aff{1}
 \corresp{\email{outi.supponen@epfl.ch}},
Danail Obreschkow\aff{2},
Marc Tinguely\aff{1},
Philippe Kobel\aff{1},
Nicolas Dorsaz\aff{1}
\and Mohamed Farhat\aff{1}}

\affiliation{\aff{1}Laboratory for Hydraulic Machines, Ecole Polytechnique F\'ed\'erale de Lausanne, Avenue de Cour 33bis, 1003 Lausanne, Switzerland
\aff{2}International Centre for Radio Astronomy Research, University of Western Australia, M468 7 Fairway, Crawley, WA 6009, Australia}

\begin{document}

\maketitle

\begin{abstract}
Fast liquid jets, called micro-jets, are produced within cavitation bubbles experiencing an aspherical collapse. 
Here we review micro-jets of different origins, scales and appearances, and propose a unified framework to describe their dynamics by using an anisotropy parameter $\zeta \geq 0$, representing a dimensionless measure of the liquid momentum at the collapse point (Kelvin impulse). 
This parameter is rigorously defined for various jet drivers, including gravity and nearby boundaries. 
Combining theoretical considerations with hundreds of high-speed visualisations of bubbles collapsing near a rigid surface, near a free surface or in variable gravity, we classify the jets into three distinct regimes: weak, intermediate and strong. 
Weak jets ($\zeta < 10^{-3}$) hardly pierce the bubble, but remain within it throughout the collapse and rebound. 
Intermediate jets ($10^{-3} < \zeta < 0.1$) pierce the opposite bubble wall close to the last collapse phase and clearly emerge during the rebound. 
Strong jets ($\zeta > 0.1$) pierce the bubble early during the collapse.
The dynamics of the jets is analysed through key observables, such as the jet impact time, jet speed, bubble displacement, bubble volume at jet impact and vapour-jet volume.
We find that, upon normalising these observables to dimensionless jet parameters, they all reduce to straightforward functions of $\zeta$, which we can reproduce numerically using potential flow theory. 
An interesting consequence of this result is that a measurement of a single observable, such as the bubble displacement, suffices to estimate any other parameter, such as the jet speed. 
Remarkably, the dimensionless parameters of intermediate and weak jets ($\zeta<0.1$) only depend on $\zeta$, not on the jet driver (i.e.~gravity or boundaries). 
In the same regime, the jet parameters are found to be well approximated by power-laws of $\zeta$, which we explain through analytical arguments. 
\end{abstract}

\section{Introduction} 
\label{s:intro}

Cavitation bubbles in liquids remain a central research topic due to their energetic properties, which can be damaging to e.g.\ hydraulic turbomachinery or ship propellers \citep{Arndt1981,Silverrad1912}, or beneficial in applications such as microfluidics \citep{Yin2005,Dijkink2008} or medicine \citep{Stride2008,Marmottant2003}.
In most cases, the damaging or beneficial effect comes from the shock and/or the micro-jet produced during the collapse of the cavitation bubbles, more specifically during the final collapse stage. 
In this paper, micro-jets always refer to the jet forming on the bubble wall and moving across the bubble interior, before piercing the wall on the opposite side.
The dynamics of these micro-jets and their diverse origins constitute the framework of this review.

Decades of detailed research revealed a remarkable diversity of behaviours and effects of micro-jets, depending on the physical conditions \citep[see reviews by][]{Blake1987,Lauterborn2010}. 
For instance, micro-jets can have diverse origins, including rigid or free surfaces near the bubble~\citep{Blake1987} or external force fields such as gravity \citep{Obreschkow2011} (section~\ref{s:origins}), and their evolution strongly depends on the properties of the liquid (section~\ref{s:quali}).
To harvest the power of jets or suppress their damaging effects, we require an understanding of their physics across all possible conditions. 
In particular, we aim for a \textit{general} description of the jet produced by a \textit{single} cavitation bubble.
Building such a general description requires both a unified theoretical model and systematic experimental studies across a wide range of parameters (e.g.\ bubble sizes, pressures, jet drivers).

Our objective is to describe the large variety of micro-jets and unify them in a single, theoretically supported framework. 
Contrary to previous works, we benefit from the luxury of increased computational power and cheaper high-speed imaging, enabling systematic numerical and experimental analyses of jetting bubbles in a large array of realistic conditions. 
Our experimental data not only cover a wide range of parameter space, but also contain some of the most spherical large cavitation bubbles and weakest jets studied to this date. 
We combine these data with selected results from the literature, covering a large diversity of jets and bubble types. 
In the aim of comparing all these data, the results are suitably normalised to a set of dimensionless parameters characterising the jet physics. 
The statistics of these parameters are then compared against systematic theoretical predictions from customised numerical simulations.
Finally, physical interpretations of the results are sought analytically.

The paper is structured as follows: 
Section~\ref{s:origins} summarises the most prominent drivers of micro-jets and quantifies their ``strength'' using a single parameter. 
Our experimental setup for the systematic investigation of jets in various conditions is described in section~\ref{s:experiment}. 
We then systematically study the variation of the micro-jet dynamics as a function of the pressure anisotropy. 
First, we phenomenologically classify the jets into three visually distinct regimes in section~\ref{s:quali}. 
Section~\ref{s:quanti} follows up with a quantitative analysis of five dimensionless jet parameters, studied as a function of a suitable anisotropy parameter and compared against numerical simulations.
Section~\ref{s:discussion} synthesises all the experimental and numerical results in a single figure, presents physical interpretations of the results and discusses potential applications and limitations.

\section{The diverse origins of micro-jets} 
\label{s:origins}

\begin{figure}
\begin{center}
\includegraphics[width=\textwidth]{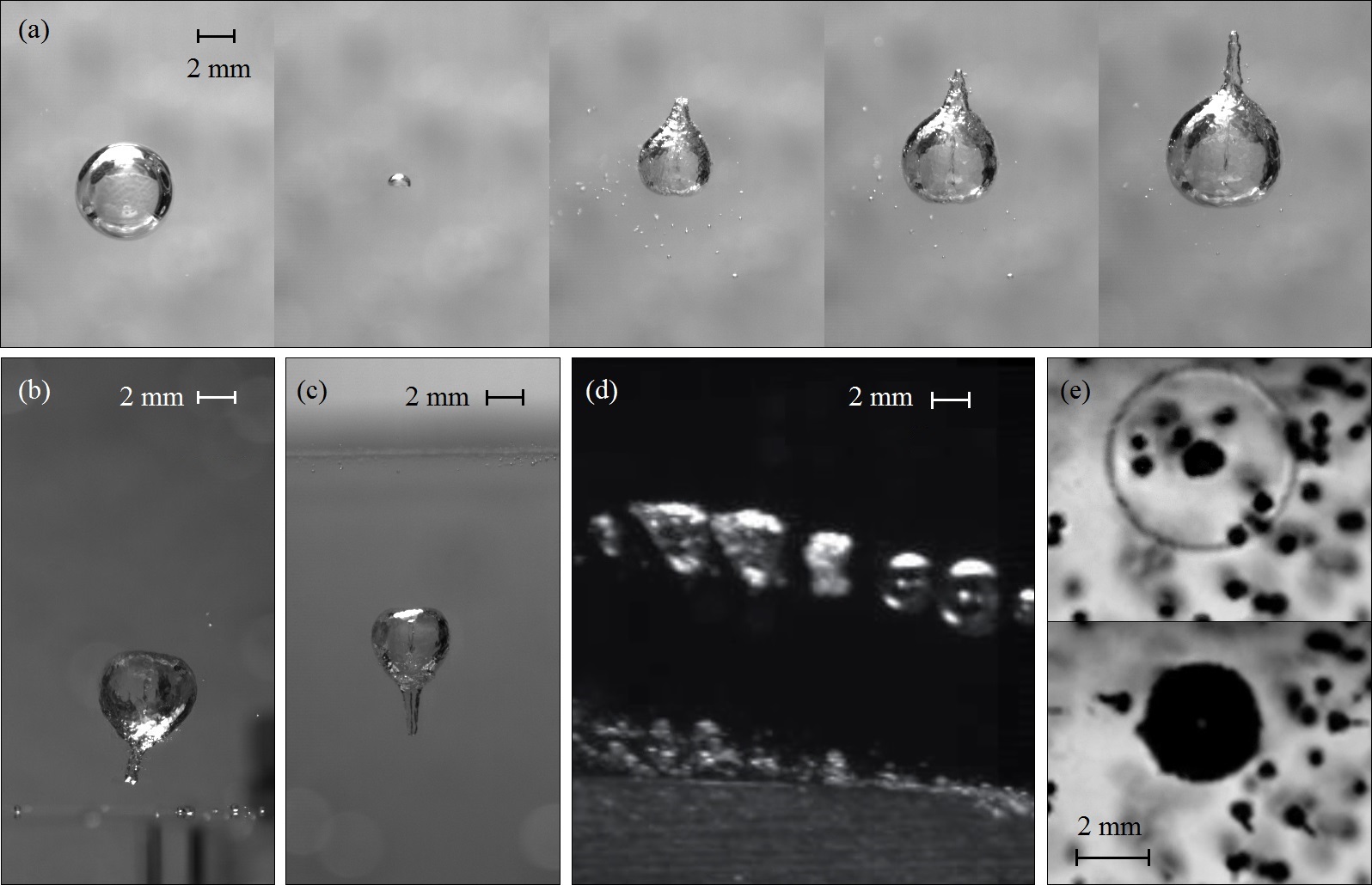}
\caption{Micro-jets from different origins: a)~Gravity, b)~rigid surface, c)~free surface, d)~stationary flow \citep[from][]{Tinguely2013} and e)~shock wave (see micro-bubbles). Images (a)-(d) correspond to equations~(\ref{eq:zeta})~(a)-(d).}
\label{fig:sources}
\end{center}
\end{figure}

Micro-jets are produced during the aspherical collapse of cavitation bubbles. 
The sphericity of bubbles is broken by anisotropies in the surrounding pressure field.
There are various possible origins for such anisotropies (see figure~\ref{fig:sources}), with the most common ones being discussed hereafter.

Most micro-jet investigations have focused on bubbles collapsing near a rigid or a free surface (figures~\ref{fig:sources}b and~\ref{fig:sources}c). 
The level of bubble asphericity is generally quantified by a dimensionless stand-off parameter $\gamma=h/R_{0}$, where $h$ is the distance from the initial bubble centre to the surface and $R_{0}$ the maximum bubble radius. 
The usual findings are that $\gamma$ governs much of the micro-jet dynamics, such as its speed or erosive force~\citep{Vogel1989,Philipp1998,Ohl1999}. Most experimental studies are limited to $\gamma<5$, as beyond this limit, the bubble undergoes a nearly spherical collapse that often appears indistinguishable from a boundary-free collapse, given the limited initial sphericity of the bubble. In highly symmetric experimental conditions (relying on mirror-focused lasers and/or microgravity conditions), it is nonetheless possible to detect jets beyond $\gamma=10$, as we shall demonstrate in sections~\ref{s:quali} and \ref{s:quanti}.

Another typical micro-jet driver, not accounted for by the stand-off parameter $\gamma$, is the gravity-induced hydrostatic pressure gradient, i.e.\ buoyancy \citep{Obreschkow2011} (figure~\ref{fig:sources}a). 
Gravity becomes particularly apparent when dealing with larger bubbles and/or hyper-gravity environments, such as in the studies by \citet{Benjamin1966}, \citet{Gibson1968} and \citet{Blake1988}. 
To quantify the effect of buoyancy, Gibson introduced the dimensionless parameter $\sigma = \rho g R_{0}/\Delta p$, where $\rho$ is the liquid density, $g$ is the gravitational acceleration and $\Delta p \equiv p_{0}-p_{v}$ is the driving pressure ($p_{0}$ is the pressure at infinity at the vertical position of the bubble centre and $p_{v}$ is the vapour pressure). 
A similar parameter, $\delta=\sigma^{1/2}$, has also been used in the past to account for the effect of gravity~\citep{Blake1988,Zhang2015}.

Further origins of cavitation bubble micro-jets are, for example, flows with pressure gradients~\citep{Tinguely2013,Blake2015} (figure~\ref{fig:sources}d), shock waves~\citep{Ohl2003,Sankin2005} (figure~\ref{fig:sources}e), focused ultrasound~\citep{Gerold2012} or neighbouring bubbles~\citep{Sankin2010}. 
Also, a combination of different jet drivers together can cause the bubble asphericity, enhancing the jet formation, or even suppressing it.
An example of such a combination is seen in figure~\ref{fig:sources}d where a bubble in a static flow collapses near a rigid hydrofoil. 
Its micro-jet, however, is not shot towards the nearest surface but, instead, directed more against the pressure gradient of the flow.

The plethora of micro-jet drivers and the fact that different drivers can act simultaneously highlights the need for a unified framework, approximately describing the jet dynamics for a multitude of jet drivers. 
To this end, we need to quantify the jet-driving pressure anisotropy with a parameter defined for various origins of this anisotropy and applicable to bubbles of many sizes and external conditions. In general, any smooth pressure field can be expanded in the space-coordinates as 
\begin{equation}
p(\mathbf{x},t_{0}) = p(\mathbf{x}_{0},t_{0}) + (\mathbf{x}-\mathbf{x}_{0})^{\top}\boldsymbol{\nabla}p + \frac{1}{2}(\mathbf{x}-\mathbf{x}_{0})^{\top}\mathsfbi{D}(p)\,(\mathbf{x}-\mathbf{x}_{0}) + O(\mathbf{x}^{3}),
\label{eq:p}
\end{equation}
where $\boldsymbol{\nabla} p$ and $\mathsfbi{D}(p)$ respectively denote the gradient and Hessian matrix of the pressure field at $\mathbf{x}=\mathbf{x}_{0}$ and $t=t_0$, here considered to be the bubble centroid and time at the instant of the bubble generation. To first order, the effects of pressure anisotropies therefore depend on the constant $\gradp$. 

To define a dimensionless anisotropy parameter, we can exploit the fact that the inviscid Navier-Stokes equations without surface tension are self-similar, such that they become dimensionless by normalising length-scales by $R_{0}$, pressures by $\Delta p$ and velocities by $(\Delta p/\rho)^{1/2}$. The assumption for the minor role of surface tension and viscosity is widely accepted for the first bubble oscillation in water for bubbles bigger than $R_{0} \sim 10^{-5}$~mm~\citep[see e.g.\ ][]{Levkovskii1968}. Applying this normalisation to $\gradp$ leads to the dimensionless vector-parameter~\citep{Obreschkow2011}
\begin{equation}
	\boldsymbol{\zeta} \equiv -\boldsymbol{\nabla} p\,R_{0}\Delta p^{-1},
\end{equation}
where the minus sign ensures that the jet driven by $\boldsymbol{\nabla} p$ is directed along $\boldsymbol{\zeta}$. A straightforward calculation (Appendix~\ref{app:derivations}) shows that $\boldsymbol{\zeta}$ is a dimensionless version of the so-called Kelvin impulse~\citep{Benjamin1966,Blake1988,Blake2015} $\mathbf{I}$, defined as the linear momentum acquired by the liquid during the asymmetric growth and collapse of the bubble,
\begin{equation}\label{eq:Igradient}
	\mathbf{I} = 4.789 R_0^3 \sqrt{\Delta p\rho}~\boldsymbol{\zeta}.
\end{equation}
The value 4.789 is strictly an irrational number, the exact value of which is given in equation~(\ref{eq:Igradientexact}) in the Appendix. The term $R_0^3 \sqrt{\Delta p\rho}$ has the units of momentum, as expected.

In situations where the micro-jet cannot be attributed to an external $\gradp$, we can define $\boldsymbol{\zeta}$ such that equation~(\ref{eq:Igradient}) still returns the correct Kelvin impulse. For instance, if the jet is caused by a rigid or free surface at a stand-off parameter $\gamma$, the Kelvin impulse is given by (Appendix~\ref{app:derivations})
\begin{equation}\label{eq:Isurface}
	\mathbf{I_{surface}} = 0.934 R_0^3 \sqrt{\Delta p\rho}~\gamma^{-2}\mathbf{n}\cdot\left\{
	\begin{array}{ll}
		-1 & {\rm flat~rigid~surface} \\
		+1 & {\rm flat~free~surface}
	\end{array}
	\right.,
\end{equation}
where $\mathbf{n}$ is the normal unit vector on the surface pointing to the cavity centre. The exact value of 0.934 is given in equation~(\ref{eq:Isurfaceexact}). Equating equations~(\ref{eq:Igradient}) and (\ref{eq:Isurface}) yields
\begin{equation}\label{eq:zeta_gamma}
	\zeta = 0.195 \gamma^{-2}.
\end{equation}
with the exact expression of 0.195 given in equation~(\ref{eq:zeta_gamma_exact}). When expressing $\zeta$ as a function of $\gamma$ in this way, equation~(\ref{eq:Igradient}) yields the correct Kelvin impulse for a rigid/free surface. An analogous approach can be used to derive $\boldsymbol{\zeta}$ for other types of boundaries~\citep{Gibson1982,Blake2015} and pressure gradients,
\begin{equation}
\boldsymbol{\zeta} = \left \{ 
 \begin{array}{l l l}
 -\rho\mathbf{g}R_{0}\Delta p^{-1} & \quad \mathrm{gravitational~field} & \quad \mathrm{(a)}\\
 -0.195\gamma^{-2} \mathbf{n} & \quad \mathrm{flat~rigid~surface} & \quad \mathrm{(b)}\\
 +0.195\gamma^{-2} \mathbf{n} & \quad \mathrm{flat~free~surface} & \quad \mathrm{(c)}\\
 -\rho(\mathbf{u}\cdot\boldsymbol{\nabla})\mathbf{u}R_{0}\Delta p^{-1} & \quad \mathrm{stationary~potential~flow} & \quad \mathrm{(d)}\\
 0.195\gamma^{-2}(\rho_{1}-\rho_{2})(\rho_{1}+\rho_{2})^{-1} \mathbf{n} & \quad \mathrm{liquid~interface} & \quad \mathrm{(e)}\\
 0.195\gamma^{-2}(4\alpha-1-8\alpha^{2}e^{2\alpha}E_{1}(2\alpha)) \mathbf{n} & \quad \mathrm{inertial~boundary} & \quad \mathrm{(f)}
 \end{array} \right.
\label{eq:zeta}
\end{equation} 
Here $\mathbf{u}$ is the velocity field, $\rho_{1}$ and $\rho_{2}$ are the different densities of two liquids, $\alpha$ is defined as $\alpha\equiv\rho h/\Sigma$ (where $\rho$ is the liquid density, $h$ is the distance from the initial bubble centre to the surface and $\Sigma$ is the surface density)~\citep{Chahine1980} and $E_1(x)\equiv\int_x^{\infty}t^{-1}e^{-t}{\rm d}t$ is an exponential integral. In the linear expansion of the pressure field, the anisotropy parameter associated with a combination of drivers (e.g.~gravity and flat surface) is given by the vector sum of the respective $\boldsymbol{\zeta}$.
Defining a corresponding anisotropy parameter for more complicated jet drivers, such as neighbouring bubbles, shock waves or ultrasound that are strongly time-dependent, or boundaries with complex geometries, is not as straightforward as for the above examples.
In the present work we focus on unifying the jet-drivers listed in equations~(\ref{eq:zeta}), and restrict experimental verification to gravity, flat rigid and free surfaces.

We expect, and will show in the following, that the jet becomes more pronounced (in a sense specified in section~\ref{s:quanti}) with increasing $\zeta \equiv |\boldsymbol{\zeta}|$. 
Importantly $\zeta$, unlike the Kelvin impulse, has the special property that bubbles with equal values of $\zeta$ produce \emph{similar} (i.e.\ identical in normalised coordinates) jets irrespective of the jet driver (e.g.\ gravity, rigid/free surface). 
This prediction naturally breaks down as the higher-order terms in equation~(\ref{eq:p}) become significant. As we shall see (section~\ref{s:quanti}), this is the case, e.g., for strongly deformed bubbles ($\zeta>0.1$, corresponding to $\gamma<1.4$ following equation~(\ref{eq:zeta_gamma})).
Following the same argument, other types of micro-jets, not treated in this work, are only well described by $\zeta$ if the time-constant gradient in the expansion of the pressure field dominates the jet formation.

\section{Experimental setup} 
\label{s:experiment}

\begin{figure}
\begin{center}
\includegraphics[width=\textwidth]{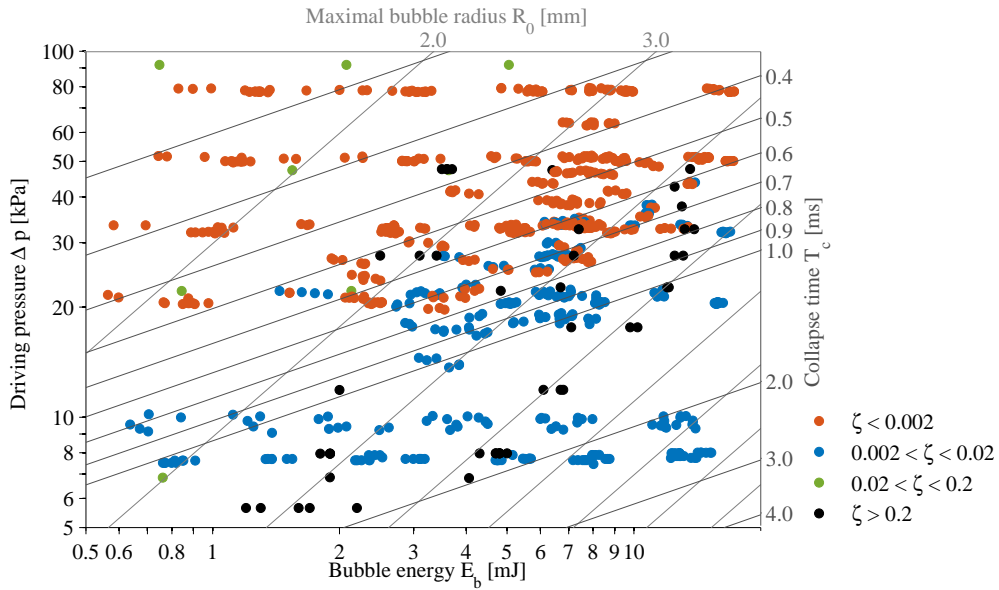}
\caption{ Overview of the parameter space covered by the experiment. The data points include bubbles subject to gravity and a nearby rigid/free surface. The four parameters $E_{b}$, $\Delta p$, $R_{0}$ and $T_{c}$ are related via the two relations $E_{b} = (4\pi/3)R_{0}^{3}\Delta p$ and $T_{c} = 0.915 R_{0}(\rho/\Delta p)^{1/2}$ (spherical collapse) and can therefore be reduced to any combination of just two parameters, representable in a two-dimensional plot.}
\label{fig:parameter}
\end{center}
\end{figure}

Our experimental setup \citep[details in][]{Obreschkow2013} generates highly spherical bubbles by focusing a green pulsed laser (532 nm, 8 ns) inside a large, cubic test chamber ($18\times18\times18$~cm$^{3}$) filled with degassed water. 
The laser beam is first expanded to a diameter of 5~cm using a lens-system, and then focused onto a single point using a parabolic mirror with a high convergence angle (53$^{\circ}$) to generate a point-like initial plasma. 
In this way, we obtain a bubble of very high initial sphericity, which is impossible to achieve with a pure lens-system that is affected by refractive index variations, spherical aberration and/or the proximity of the lens to the bubble. 
As a result, we are able to cover a large range of anisotropies $\zeta$, including the delicate ``weak jet'' regime previously unexplored, where the jets are barely observable (see section~\ref{s:weak}). 
We observe the micro-jets through high-speed visualisations with the Photron SA1.1 and Shimadzu HPV-X camera systems, reaching speeds up to 10~million frames per second. The bubbles are illuminated using a flash-lamp (bubble interface and interior) or a parallel backlight LED (shadowgraphy and shock waves).

Three parameters can be independently varied in our experiment: (i)~the driving pressure $\Delta p$ ($\sim 0.1$-1~bar), (ii)~the bubble energy $E_{b} = (4\pi/3)R_{0}^{3}\Delta p$ (1-12~mJ) and (iii)~the gravity-induced pressure gradient $\boldsymbol{\nabla} p$, modulated aboard ESA parabolic flights (56th, 60th and 62nd parabolic flight campaigns).
In addition, a free or a rigid surface may be introduced near the bubble at a controlled distance.
The maximum bubble radii $R_{0}$ vary within the range $1.5-8.0$~mm and the Rayleigh collapse times ($T_{\rm c} = 0.915 R_{0}(\rho/\Delta p)^{1/2}$) within the range $0.1-3$~ms. 
The parameter space covered by the experiment is displayed in figure~\ref{fig:parameter}. A sub-sample of these data points is used in the following analyses.

\section{Qualitative classification of jetting regimes} 
\label{s:quali}

\begin{figure}
\begin{center}
\includegraphics[width=.8\textwidth]{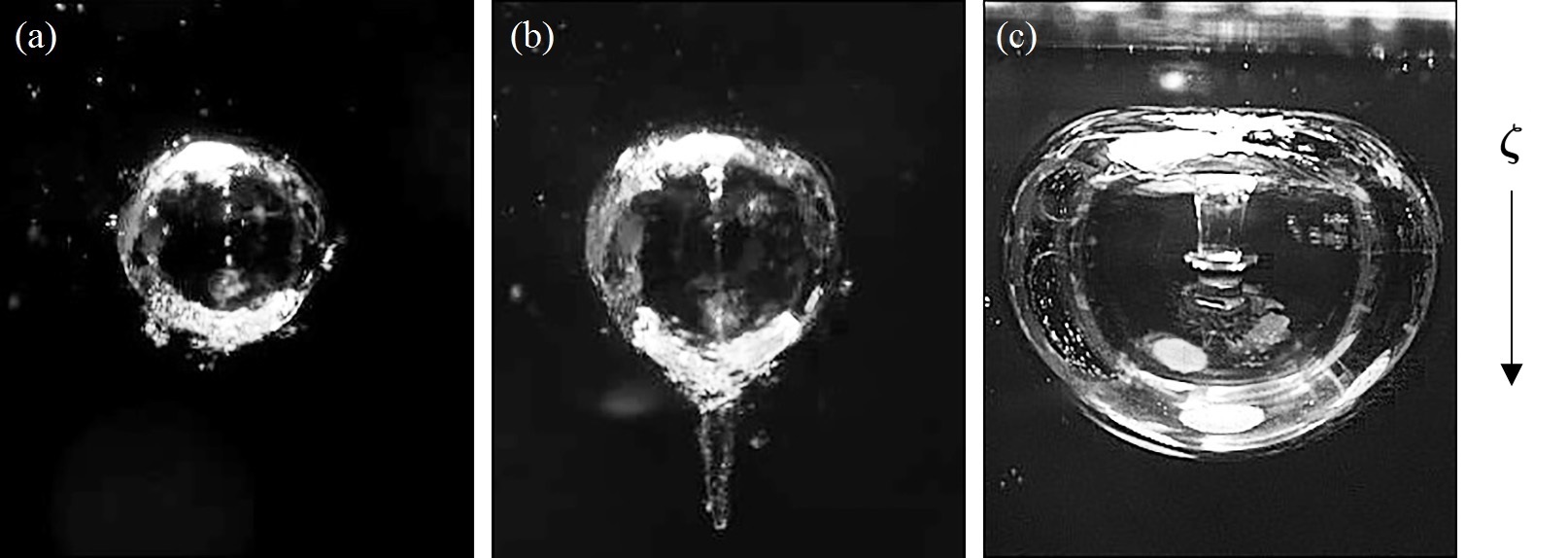}
\caption{Observations of three distinct micro-jet types driven by a nearby free surface: (a)~weak jet ($\zeta \lesssim 0.001$) seen only inside the rebound bubble following the collapse, (b)~intermediate jet ($\zeta = 0.01$) emerging during the rebound and (c)~strong jet \citep[$\zeta = 0.64$, from][]{Supponen2015} seen early during the collapse. The arrow on the right shows the direction of the anisotropy parameter $\boldsymbol{\zeta}$.}
\label{fig:regimes}
\end{center}
\end{figure}

The micro-jet dynamics strongly varies with the anisotropy in the pressure field, that is with the anisotropy parameter $\zeta$ defined in eq.~(\ref{eq:zeta}). This section introduces a phenomenological classification of the micro-jets dynamics into three separate regimes, ``weak'', ``intermediate'' and ``strong'', identified with three distinct ranges of $\zeta$. 
An example of a micro-jet in each regime is given in figure~\ref{fig:regimes}: Weak (a) and intermediate (b) jets form so close to the collapse point that they are primarily visible during the rebound. 
Whereas intermediate jets push through the wall of the rebound bubble and drag along a conical vapour pocket (``vapour-jet''), weak jets hardly pierce the rebound bubble and remain almost entirely inside it. In turn, strong jets (c) pierce the bubble well before the first collapse, leaving behind thick vortex rings. 

The transition between weak and intermediate jets occurs around $\zeta=10^{-3}$, whereas the division between intermediate and strong jets lies around $\zeta=0.1$. These transitions are not sharp, since the jet dynamics changes continuously with $\zeta$. The separation between weak, intermediate and strong jets nonetheless presents a useful thinking tool to establish a unified perspective on these visually distinct types of micro-jets. Each regime is discussed and visualised in detail in the following sections.

\subsection{Weak jets ($\zeta\leq10^{-3}$)} 
\label{s:weak}

\begin{figure}
\begin{center}
\includegraphics[width=\textwidth]{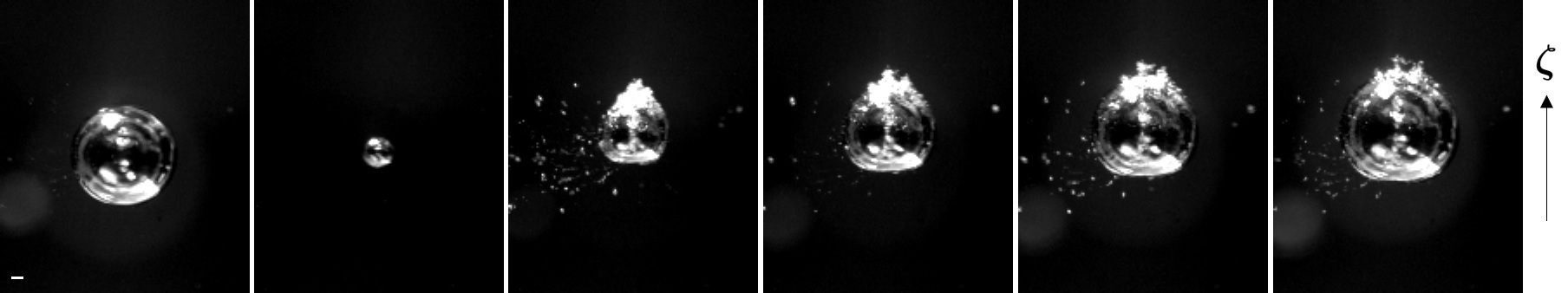}
\caption{Weak jet formation driven by gravity. The interframe time is 90 $\mu$s. The white bar shows the 1~mm scale. The anisotropy parameter $\zeta$ equals 0.001. The arrow on the right shows the direction of $\boldsymbol{\zeta}$. See \textit{Movie1.mp4}.}
\label{fig:weak}
\end{center}
\end{figure}
\begin{figure}
\begin{center}
\includegraphics[width=\textwidth]{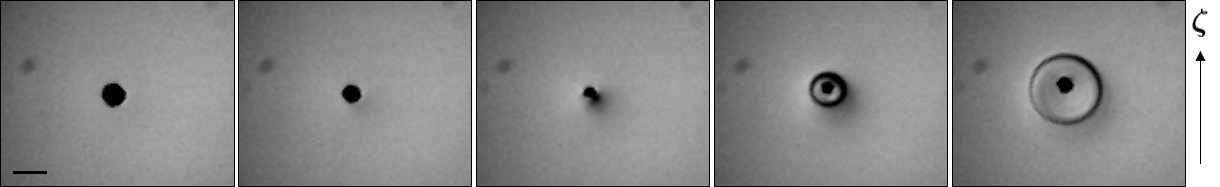}
\caption{Shock wave emission at the collapse of a bubble with a gravity-driven weak jet. The interframe time is 300~ns. The exposure time is only 60~ns, leading to a sharp shock front. The black bar shows the 1~mm scale. The anisotropy parameter $\zeta$ equals 0.001. The arrow on the right shows the direction of $\boldsymbol{\zeta}$. See \textit{Movie2.mp4}.}
\label{fig:weakshock}
\end{center}
\end{figure}

Weak jets are the most delicate type of micro-jets. They are only seen during the rebound phase succeeding the first bubble collapse, and even then they remain entirely, or almost entirely, contained \emph{inside} the rebound bubble. Therefore, weak jets can only be revealed using sophisticated visualisations of the bubble interior.

The reason why weak jets merit a regime of their own, despite their hidden existence, is the sensitivity of the collapse physics on even tiny pressure anisotropies.
For instance, the luminescence energy of bubbles near boundaries has been shown to vary with the stand-off parameter $\gamma$ up to $\gamma \approx 20$~\citep{Ohl1998} ($\zeta \approx 5\cdot10^{-4}$). 
We find this to be the case for even lower values of $\zeta$ (discussed in a forthcoming publication).

Experimentally, an extremely high initial bubble sphericity is required for a weak jet to form.
Based on numerical models used to design the experimental setup (section \ref{s:experiment}), we estimate that the amplitude of the deformation of the initial bubble relative to its maximal radius should be less than $10^{-4}$. 
Bubbles generated by discharge-sparks~\citep[e.g.][]{Gibson1968} and lens-focused laser pulses~\citep[e.g.][]{Philipp1998} are generally not spherical enough to probe the regime $\zeta < 10^{-2}$ \citep[see][chapter 4]{Tinguely2013}. 
Within the accuracy of such standard experiments, $\gamma > 4$ (or $\zeta < 0.012$) appears to produce a spherical collapse~\citep{Isselin1998}, where, in fact, the jet has been masked by perturbations that are more important than the jet itself. 
The hidden weak jet is also challenging to visualise due to its microscopic size, its unstable nature within the rebound and a non-transparency of the bubble interface at the early rebound stages. 

Our experiment (section \ref{s:experiment}) is suitable for studying weak jets by virtue of its mirror-focused laser and the option to reduce gravity on parabolic flights. 
An example of a weak jet produced by a distant free surface ($\gamma\approx14$) is shown in figure~\ref{fig:regimes}a. 
An alternative example of a gravity-driven weak jet is shown in figure~\ref{fig:weak}.
The bubble remains highly spherical throughout the collapse (frames 1--2) and rebound (frames 3--6). 
However, one can observe a jet inside the rebound bubble (frames 3--4). 
During the growth of the rebound bubble, the micro-jet becomes unstable and `pulverises' into a chain of microscopic droplets. (The phenomenon is more readily observable in the linked video.)

Bubbles with weak jets emit a single shock at their collapse, as shown in figure~\ref{fig:weakshock}. 
The only way to tell that the bubble is subject to a deformation during its collapse is its translation, which is an expression of the momentum (Kelvin impulse) accumulated during the growth and collapse. 
The bubble has moved most significantly at its minimal radius between frames 3 and 4 in figure~\ref{fig:weakshock}, as evidenced by the different centres of the bubble and the shock in frame 4.

By systematically varying $\zeta$ while taking visualisations similar to figure~\ref{fig:weak}, we found $\zeta\leq10^{-3}$ (corresponding to $\gamma\gtrsim14$ for bubbles near a rigid or free surface) to be the anisotropy range of weak jets. 
Larger values of $\zeta$ produce jets that visibly emerge from the rebound bubble (see section~\ref{s:intermediate}). 
The limit is not a hard one, but nonetheless gives a fair indication on the pressure anisotropy where a significant reduction in the vapour-jet size outside the rebound bubble is observed.

The observed instability of weak jets, as well as the fact that these jets live entirely inside the bubble gas -- a medium of rapidly changing temperature and pressure -- hint at complex physical mechanisms, beyond the scope of this work. A subtle question is whether a weak jet slightly pierces the bubble at the collapse point. Potential flow theory of an empty bubble predicts that the jet always pierces the bubble~\citep{Blake1987} no matter how small the Kelvin impulse ($>0$). However, our visualisations do not show clear evidence for such piercing -- at least the jet does not entrain a vapour-jet. Perhaps, weak jets are so small and low in kinetic energy that they are stopped by surface tension or heavily affected by the hot plasma at the last collapse stage. Detailed modelling, ideally using molecular dynamics simulations, is needed to uncover these details.

\subsection{Intermediate jets ($10^{-3}<\zeta<0.1$)} 
\label{s:intermediate}

In the intermediate jet regime ($10^{-3} < \zeta < 0.1$), the jet pierces the bubble close to the moment of collapse and entrains a conical vapour-jet during the rebound phase.

\begin{figure}
\begin{center}
\includegraphics[width=\textwidth]{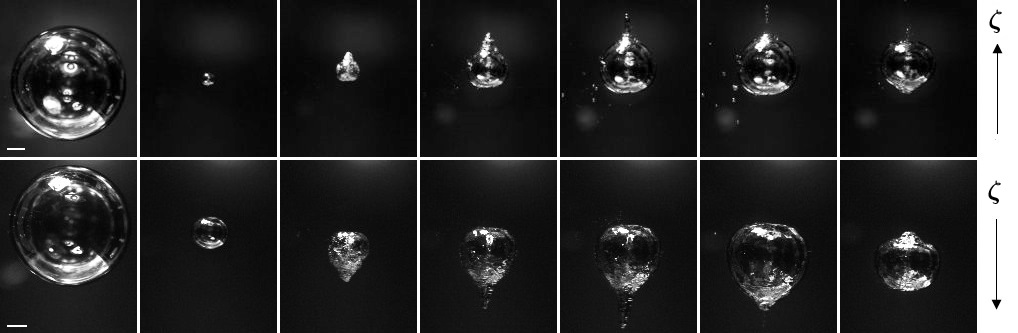}
\caption{Selected images of bubbles with intermediate jets driven by gravity (upper) and a nearby free surface (lower). Images have been taken at times $t =$ 0.9, 2.15, 2.25, 2.35, 2.45, 2.85 and 3.35 ms (upper) and $t =$ 2.05, 4.15, 4.2, 4.35, 4.6, 4.75 and 6.2 ms (lower) from bubble generation. (The different evolution speeds are simply due to different liquid pressures chosen for the two experiments.) The white bar shows the 1~mm scale. The anisotropy parameter $\zeta$ equals 0.007, equivalent to a stand-off parameter $\gamma$ of 5.3. The arrows on the right show the direction of $\boldsymbol{\zeta}$. See \textit{Movie3.mp4} and \textit{Movie4.mp4}.}
\label{fig:intermediate}
\end{center}
\end{figure}

Figure~\ref{fig:intermediate} shows an intermediate jet produced by gravity (upper) and by a nearby free surface (lower). 
The jet is visible inside the rebound bubble and as a conical protrusion of vapour dragged along while the jet is penetrating the liquid. 
The rebound bubble has a transparent interface and eventually regains a shape close to spherical. It is worth emphasising that, despite the different jet drivers in figure~\ref{fig:intermediate}, both bubbles exhibit nearly identical shapes apart from the opposite jet directions. This confirms our expectation (section~\ref{s:origins}) that identical values of $\zeta$ lead to similar jets, independently of the jet driver.

One can note a similar pulverisation of the jet inside the rebound bubble as observed in the case of the weak jets (more readily visible in the linked video). 
Furthermore, the issue of initial bubble sphericity discussed in section~\ref{s:weak} plays an important role in the intermediate regime as well. 
Micro-jet studies in the literature seldom observe jets at $\gamma > 4$~\citep{Isselin1998}, while we observe both gravity- and boundary-induced jets all the way down to the weak jet regime at $\zeta<10^{-3}$, corresponding to $\gamma>14$.

\begin{figure}
\begin{center}
\includegraphics[width=\textwidth]{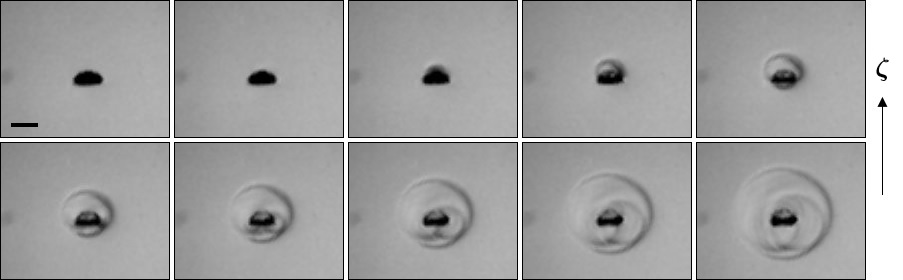}
\caption{Shock wave emission at the collapse of a bubble with a gravity-driven intermediate jet. The interframe time is 100~ns, the exposure time is 60~ns. The black bar shows the 1~mm scale. The anisotropy parameter $\zeta$ equals 0.007. The arrow on the right shows the direction of $\boldsymbol{\zeta}$. See \textit{Movie5.mp4}.}
\label{fig:intermediateshock}
\end{center}
\end{figure}

There is a peculiarity that we observe in the intermediate regime: the formation of a bump on the rebound bubble, at the location where the micro-jet initially develops (i.e.\ opposite from where the jet pierces the bubble). This bump can be seen in the last frames of figure~\ref{fig:intermediate}. \cite{Vogel1989} explained this phenomenon as a wake of a vortex ring inside the bubble, induced by the ring vortex in the liquid surrounding the rebounding bubble. 
However, our visualisations suggest that it is the pinch-off and the break-up of the jet within the rebound bubble that cause this deformation. 
Due to surface tension, the remainder of the jet is pulled back and seen as a bulge on the interface. 
This part of the interface struggles to follow the rest of the bubble during the second collapse, making the deformation even more pronounced (see linked videos in figure~\ref{fig:intermediate}). 
Such a deformation is predominantly seen in bubbles collapsing in the intermediate regime, although it is also marginally observed in bubbles with weak jets.

In the intermediate regime, the piercing of the bubble occurs so late in its lifetime that extreme temporal and spatial resolutions are needed to capture the jet before the collapse point. Interestingly, shock wave visualisations can be exploited to increase the time-resolution much beyond the frame-rate by virtue of the high shock velocities. The multiple shock waves in figure~\ref{fig:intermediateshock}, in particular the different radii of these shocks, clearly reveal that the jet pierces the bubble before the collapse of the torus, even though this is hard to see by looking at the bubble itself. 

\begin{figure}
\begin{center}
\includegraphics[width=\textwidth]{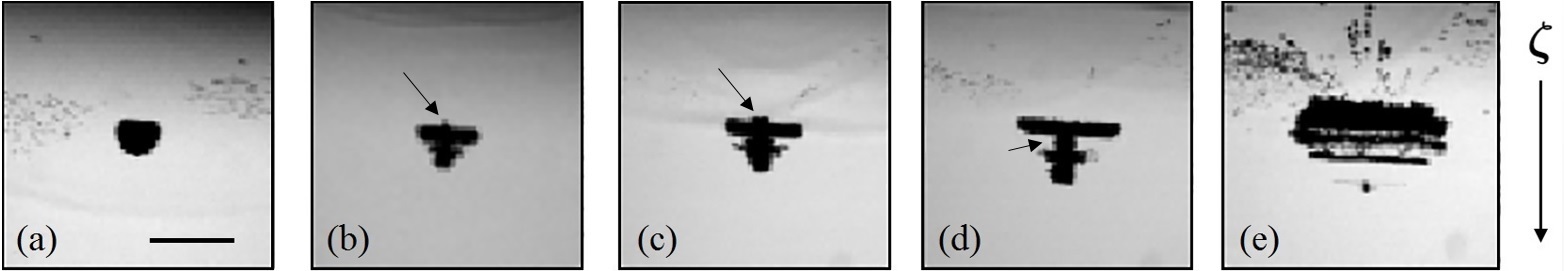}
\caption{Selected images right after the collapse of bubbles near a free surface with (a)~$\gamma =$~2.1, (b)~$\gamma =$~1.6, (c)~$\gamma =$~1.3, (d)~$\gamma =$~1.0 and (e)~$\gamma =$~0.86. Counter-jet formation is visible in (b), (c) and (d), indicated with arrows. The black bar shows the 1~mm scale. The arrow on the right shows the direction of $\boldsymbol{\zeta}$. }
\label{fig:counterjet}
\end{center}
\end{figure}

An interesting feature that many micro-jet studies have come across in the intermediate jet regime (and partly in the strong jet regime) is the apparition of a ``counter-jet'' that appears right after the bubble collapse and moves in the opposite direction to the original micro-jet. Such a counter-jet has been reported to appear for bubbles collapsing near rigid surfaces at $1<\gamma<3$~\citep{Lindau2003} and to consist of a cluster of tiny bubbles. The formation of the counter-jet is attributed to the jet impact on the opposite bubble wall. However, the phenomenon has also been seen in bubbles with gravity-driven jets at $\zeta \approx 0.2$ \citep[see][figure 2]{Zhang2015}. Furthermore, in our experiment we observe such counter-jets for bubbles collapsing near a free surface, as seen in figure~\ref{fig:counterjet} (visible in (b) and (c), also in (d) -- although here the counter-jet does not appear above the torus but rather as a ``column'' on the central axis of the torus). The phenomenon is therefore not linked to the presence of rigid boundaries but to the pressure anisotropy of the aspherical collapse. The formation of the counter-jet has been suggested to be a result of the self-penetration of the ``jet torus-shock waves'', i.e.\ the shock waves emitted at the collapse of the main torus, that create a region of tension perpendicular to the torus ring at their confluence~\citep{Lindau2003}.

\subsection{Strong jets ($\zeta>0.1$)} 
\label{s:strong}

The strong jet regime ($\zeta > 0.1$) is characterised by the jet piercing the bubble well before (more than 1\%, cf.\ section~\ref{s:jetimpact}) the collapse. 
Strong jets have mostly been observed near a rigid or a free surface~\citep{Philipp1998,Zhang2013}, but also gravity has been shown to produce jets in this regime~\citep{Zhang2015}.

\begin{figure}
\begin{center}
\includegraphics[width=\textwidth]{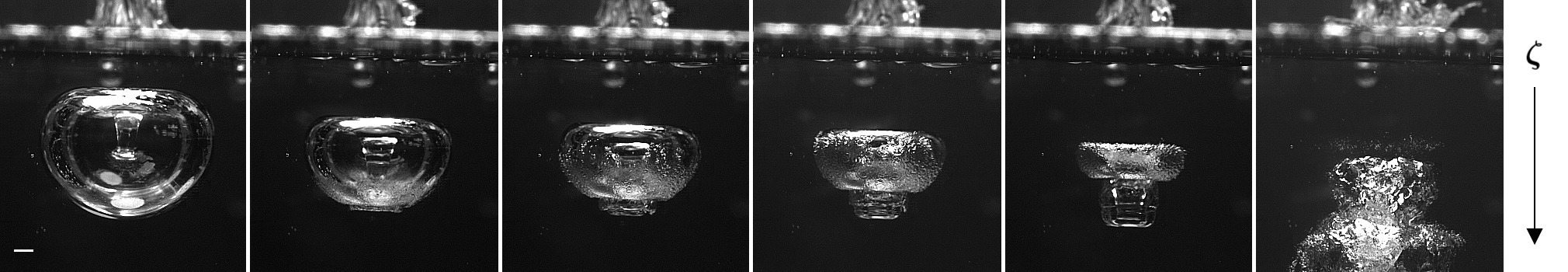}
\caption{Selected images of a bubble with a strong jet driven by a nearby free surface \citep[from][]{Supponen2015}. The anisotropy parameter $\zeta$ equals 0.62, equivalent to a stand-off parameter $\gamma=0.56$. The white bar shows the 1~mm scale. The arrow on the right shows the direction of $\boldsymbol{\zeta}$. Video: APS-DFD (\url{http://dx.doi.org/10.1103/APS.DFD.2014.GFM.V0084})}
\label{fig:strong}
\end{center}
\end{figure}

The strong jet regime is peculiar in the sense that the complex collapse dynamics involved is highly sensitive to the origin of the pressure anisotropy.
For instance, there is a large variety in shapes the jet can take prior to piercing the bubble, from large and broad \citep[such as in figure 3 in][]{Zhang2015} to thin, mushroom-capped jets~\citep{Supponen2015} typically linked to a nearby free surface (such as in figure~\ref{fig:strong}). 

The collapse of a strongly jetting bubble follows a sequence of highly complex dynamics. 
Figure~\ref{fig:strong} shows an example of such a bubble collapsing near a free surface ($\zeta$ = 0.64, i.e.\ $\gamma$ = 0.56), the micro-jet being particularly thin compared to the bubble size. 
The interface of the bubble becomes opaque already prior to the collapse (frames 3-5) due to perturbations caused by the jet impact on the opposite side of the bubble~\citep{Supponen2015}. 
Following the jet impact, the bubble breaks into two parts as a vapour pocket is entrained by the jet. 
Each part has its individual collapse. 
The rebounding bubble emerges as a chaotic bubble cloud (frame 6). 

\begin{figure}
\begin{center}
\includegraphics[width=\textwidth]{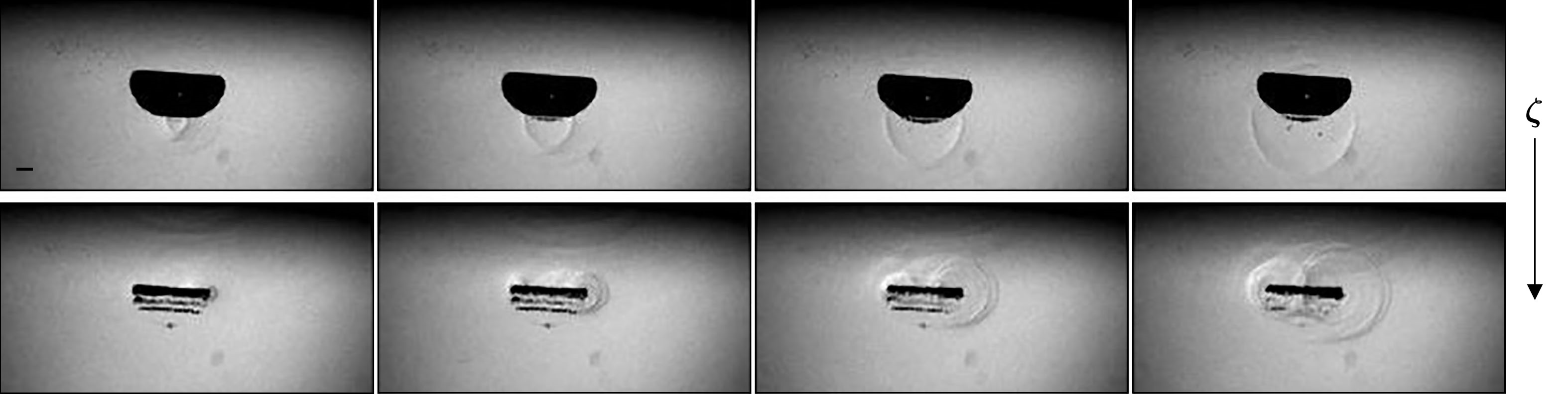}
\caption{Shock wave emission at the collapse of a bubble with a free surface-driven strong jet, with the jet impact (upper) and toroidal collapse (lower) \citep[from][]{Supponen2015}. The interframe time is 300~ns, the exposure time is 60~ns. The black bar shows the 1~mm scale. The anisotropy parameter $\zeta$ equals 0.22, equivalent to a stand-off parameter $\gamma=0.95$. The arrow on the right shows the direction of $\boldsymbol{\zeta}$. Video: APS-DFD (\url{http://dx.doi.org/10.1103/APS.DFD.2014.GFM.V0084})}
\label{fig:strongshock}
\end{center}
\end{figure}

Figure~\ref{fig:strongshock} displays a shock wave visualisation of another strongly jetting bubble collapsing near a free surface (at lower $\zeta$). 
A first shock wave is emitted at the jet impact on the bubble wall (upper row), and a complex pattern of shock waves is generated as the bubble breaks down into different tori that each collapse individually~\citep{Lauterborn1997,Supponen2015}.

Important variations for different jet drivers (gravity versus rigid/free surfaces) are expected at these high pressure anisotropies, as a direct consequence of the higher-order terms in eq.~(\ref{eq:p}). These higher-order terms and their time-dependence ensure that a bubble next to a rigid boundary ($\gamma<1$) cannot cross that boundary and that a bubble next to a free surface ($\gamma<0.5$) will burst that surface, while bubbles with a comparable Kelvin impulse generated by gravity simply travel large distances ($>R_0$, see section \ref{s:quanti}). 


\section{Quantitative analysis of jet dynamics} 
\label{s:quanti}

We now present different quantitative parameters describing micro-jets across all three jetting regimes of section \ref{s:quali}. 
We complement our experimental results with selected data from the literature for the following jet types: gravity-induced, free surface-induced and rigid surface-induced micro-jets, as well as combinations thereof.
These data also cover a large diversity of bubble types, including bubbles generated by pulsed lasers (with lens and mirror focus), sparks, underwater explosions and focused ultrasound. 

The experimental data are compared against theoretical models based on potential flow theory. 
We start the section by presenting these numerical models, and subsequently discuss how the normalised jet impact timing, the jet speed, the bubble centroid displacement, the bubble volume at jet impact and the vapour-jet volume vary with the pressure anisotropy, quantified by $\zeta$.

\subsection{Numerical simulation} 
\label{s:num}

\begin{figure}
\begin{center}
\begin{overpic}[width=0.9\textwidth]{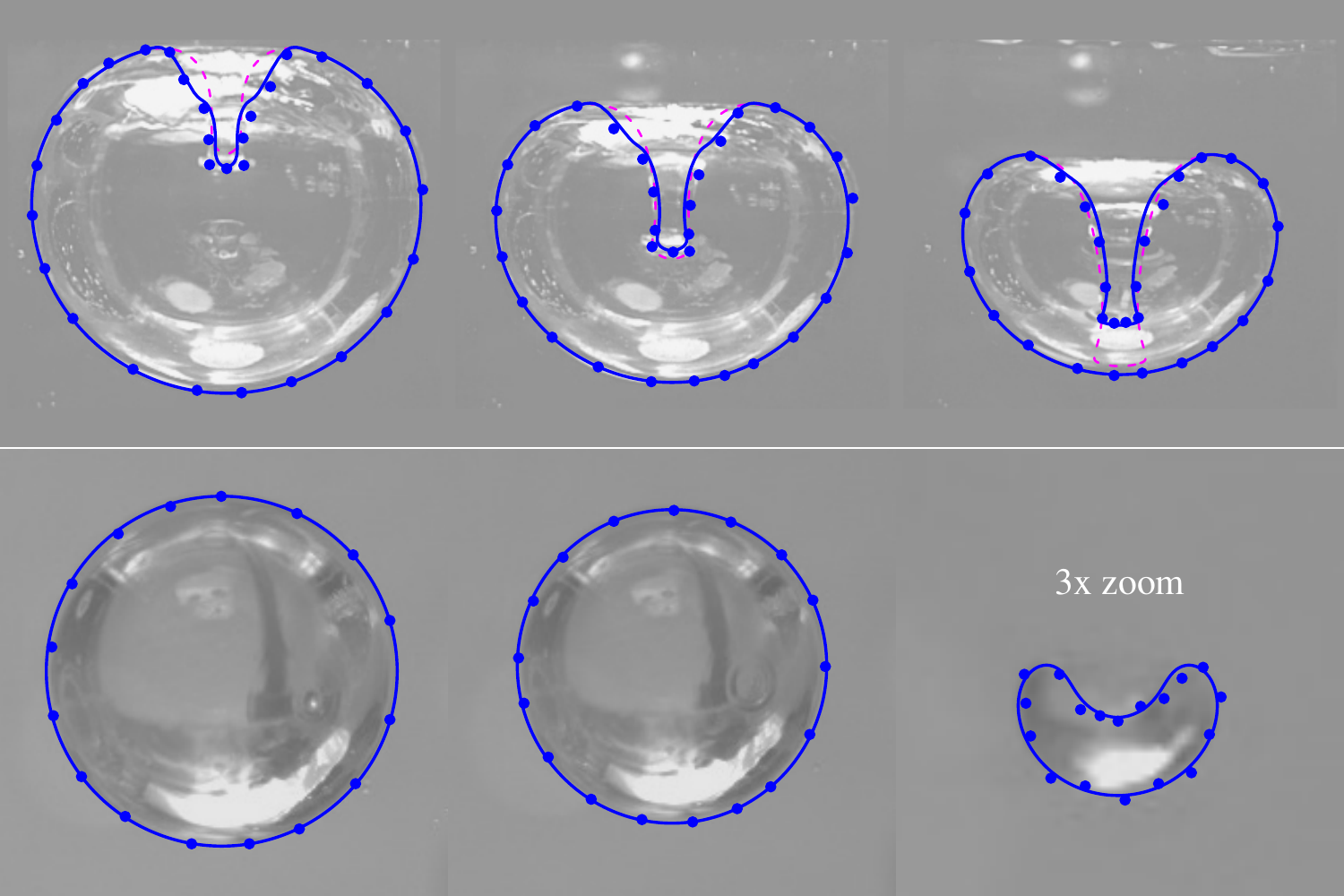}
\end{overpic}
\caption{The numerical simulations superimposed with the experimental visualisations for a bubble collapsing near a free surface $\gamma = 0.56$ (top) and near a rigid surface $\gamma = 2.32$ (bottom).
The blue points are extracted from the observed bubble shapes and the lines represent simulated data. In the case of the upper panel, the simulated bubble shape (dashed purple line) was corrected for optical refraction (solid blue line) by the outer bubble boundary, assuming a refraction by a sphere with equations analogous to those in~\cite{Kobel2009} (with water and vacuum inverted).}
\label{fig:comparison}
\end{center}
\end{figure}

We calculate the evolution of the bubble and the formation of the micro-jet in the standard model of an inviscid, incompressible fluid without surface tension. 
The bubble is assumed to contain fully condensable gas of constant pressure $p_v$. The pressure infinitely far away from the bubble, at the vertical level of the bubble centroid, is $p_0$. The evolution of this bubble is governed by the simplified Navier-Stokes equations
\begin{eqnarray}
 \frac{\mathrm{D}\mathbf{u}}{\mathrm{D}t} & = & -\boldsymbol{\nabla}p/\rho+\mathbf{g}\label{eq:numerical1},\\
 \boldsymbol{\nabla} \cdot \mathbf{u} & = & 0\label{eq:numerical2},
\end{eqnarray}
where $D\mathbf{u}/\mathrm{D}t\equiv\partial\mathbf{u}/\partial{t}+(\mathbf{u}\cdot\boldsymbol{\nabla})\mathbf{u}$ is the material derivative, i.e.\ the time-derivative seen by a particle moving with the flow. Equations (\ref{eq:numerical1}) and (\ref{eq:numerical2}) represent the conservations of momentum and mass, respectively. These equations must be completed with suitable initial and boundary conditions that depend on the jet driver (e.g.\ rigid surface~\citep{Taib1983}, free surface or gravity~\citep{Robinson2001}).

\begin{figure}
\begin{center}
\includegraphics[width=\textwidth]{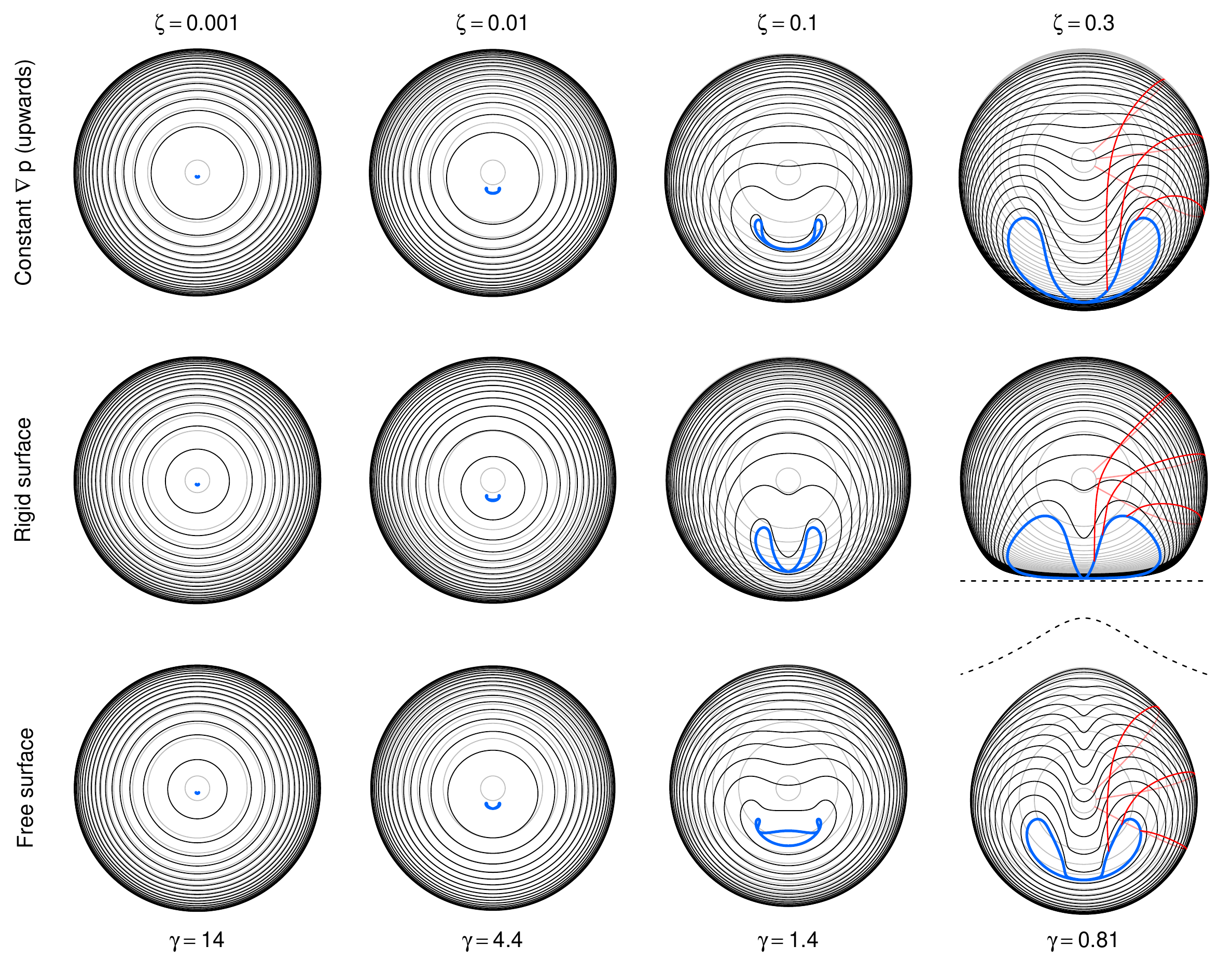}
\caption{Calculated examples of bubbles collapsing in a constant pressure gradient (upper), near a rigid surface (middle) and near a free surface (lower) at corresponding pressure anisotropy $\zeta$ and stand-off $\gamma$. The bubble shapes are shown during its growth (grey), collapse (black) and jet impact stage (blue). Surface particle trajectories are shown in red for the bubbles at $\zeta=0.3$. The dashed lines represent the rigid/free surface.}
\label{fig:simu}
\end{center}
\end{figure}

A straightforward, but numerically delicate method for solving these equations is the ``pressure formulation'', where eq.~(\ref{eq:numerical2}) is rewritten as a condition on the time-dependent pressure field $p$ needed to evaluate $\boldsymbol{\nabla} p$ in eq.~(\ref{eq:numerical1}). A more powerful and precise method, strongly advocated by Blake and collaborators~\citep{Taib1983,Robinson2001,Blake1987}, is the boundary integral method. This method relies on the flow being irrotational, $\boldsymbol{\nabla}\times\mathbf{u}=0$, such that the velocity field $\mathbf{u}$ derives from a potential $\phi$, via $\mathbf{u} = \boldsymbol{\nabla}\phi$. Green's integral formula~\citep{Blake1987} applied to eq.~(\ref{eq:numerical2}) then leads to 
\begin{equation}
\phi(\mathbf{r})= \frac{1}{2\pi}\left[\int_{\mathbf{r'} \in S}\mathrm{d}S\frac{\partial \phi(\mathbf{r'})}{\partial n}\frac{1}{|\mathbf{r}-\mathbf{r'}|} - \int_{\mathbf{r'} \in S}\mathrm{d}S\,\phi(\mathbf{r'})\frac{\partial}{\partial n}\left(\frac{1}{|\mathbf{r}-\mathbf{r'}|}\right)\right],
\label{eq:phi1}
\end{equation}
where $S$ denotes the surface of the bubble and, if present, the free surface of the liquid; and $\partial/\partial n$ denotes the normal derivative on that surface away from the liquid.

The time-evolution of the potential is given by Bernoulli's principle, which derives from eq.~(\ref{eq:numerical1})~\citep{Robinson2001,Taib1983},
\begin{equation}
\frac{\mathrm{D}\phi}{\mathrm{D}t} = \frac{|\mathbf{u}|^{2}}{2} - g z + P
\label{eq:phi2}
\end{equation}
\noindent where $z$ denotes the direction against the gravity vector $\mathbf{g}$, $g$ is the norm of $\mathbf{g}$, and the pressure term is given by $P=\Delta p/\rho=(p_v-p_0)/\rho$ on the bubble surface and $P=0$ on the free surface.

We discretise and numerically solve equations (\ref{eq:phi1}) and (\ref{eq:phi2}) using the scheme presented in~\cite{Taib1983}. This method discretises the boundary into linear elements in which case equation~(\ref{eq:phi1}) can be rewritten as a linear system of equations. It should be noted that the model only computes the bubble evolution up to the moment of jet impact, i.e.\ when the bubble becomes toroidal.

A crucial feature of the model specified by equations (\ref{eq:phi1}) and (\ref{eq:phi2}) is that, upon normalising distances to the maximal bubble radius $R_0$ and normalising the time to $R_0(\rho/\Delta p)^{1/2}$, the evolution of the bubble exclusively depends on the anisotropy parameter $\zeta$ given in equation~(\ref{eq:zeta}) and on the origin of $\zeta$ (e.g.\ gravity or nearby surfaces) via the boundary conditions. Moreover, since $\zeta$ is defined such that to first order the pressure anisotropy does \emph{not} depend on the origin, we expect the micro-jet to depend on the origin only for large values of $\zeta$.

The bubble shapes calculated through the numerical simulation are superimposed with the corresponding experimental images in figure~\ref{fig:comparison} with two distinct jet drivers.
The simulated and observed shapes are in good agreement, justifying the use of the boundary integral method for the analysis of the individual micro-jet parameters.
Interestingly, even the ``mushroom-cap''-shaped jet tip is reproduced for the bubble collapsing near a free surface (note the optical distortion of the jet tip in the final image).

The simulation neglects viscosity and surface tension, which could have an effect on the detailed jet shape.
Nevertheless, these should have a minor role to the total Kelvin impulse, most of which is accumulated when the jet is in its early formation stage.
We also note that the boundary integral method does not fully satisfy the no-slip condition, potentially important when the bubble is very close to a rigid surface.

\begin{figure}
\begin{center}
\includegraphics[width=\textwidth]{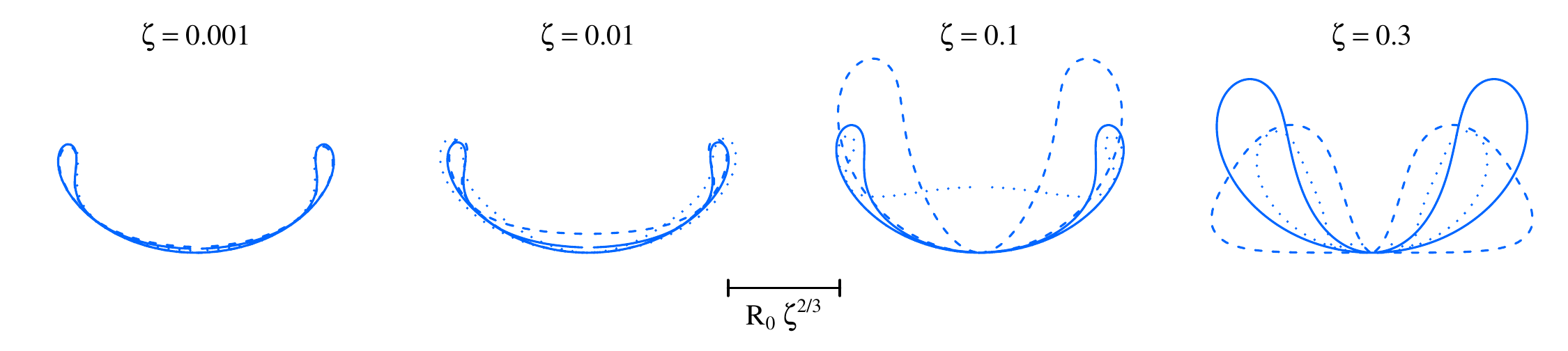}
\caption{Zoomed bubble shapes at the jet impact from figure~\ref{fig:simu}. The different jet drivers are indicated by solid (constant $\gradp$,) dashed (rigid surface) and dotted (free surface) lines. The scale bar shows the characteristic scale of the final bubble as explained in section~\ref{s:interp}.}
\label{fig:zoom}
\end{center}
\end{figure}

Figure~\ref{fig:simu} displays examples of calculated bubble shapes at different levels of $\zeta$ (and corresponding $\gamma$, related to $\zeta$ via equation~(\ref{eq:zeta_gamma})), across all regimes ($\zeta = 0.001$ is the limit between weak and intermediate jet regimes, $\zeta=0.01$ is in the intermediate jet regime, $\zeta=0.1$ is the limit between intermediate and strong jet regimes and $\zeta = 0.3$ is in the strong jet regime).
This figure illustrates the differences of a bubble collapsing in a constant pressure gradient, near a rigid surface and near a free surface. 
The differences in the bubble shapes are significantly more pronounced in the strong jet regime compared to the weak and intermediate jet regimes.
We show this explicitely by zooming into the bubble shapes at the instant of the jet impact in figure~\ref{fig:zoom}.
One should therefore expect important differences in the quantitative properties of micro-jets in the strong jet regime. 
In turn, in the intermediate and weak jet regimes the micro-jets are well described by $\zeta$, independently of the origin of the anisotropy. 
We will verify this statement by looking at individual micro-jet parameters in the following sections.

The code used to solve equations (\ref{eq:phi1}) and (\ref{eq:phi2}) is available online at \url{https://obreschkow.shinyapps.io/bubbles}. 

\subsection{Jet impact time} 
\label{s:jetimpact}

\begin{figure}
\begin{center}
\includegraphics[width=\textwidth, trim=1.9cm 0cm 3cm -0.3cm, clip]{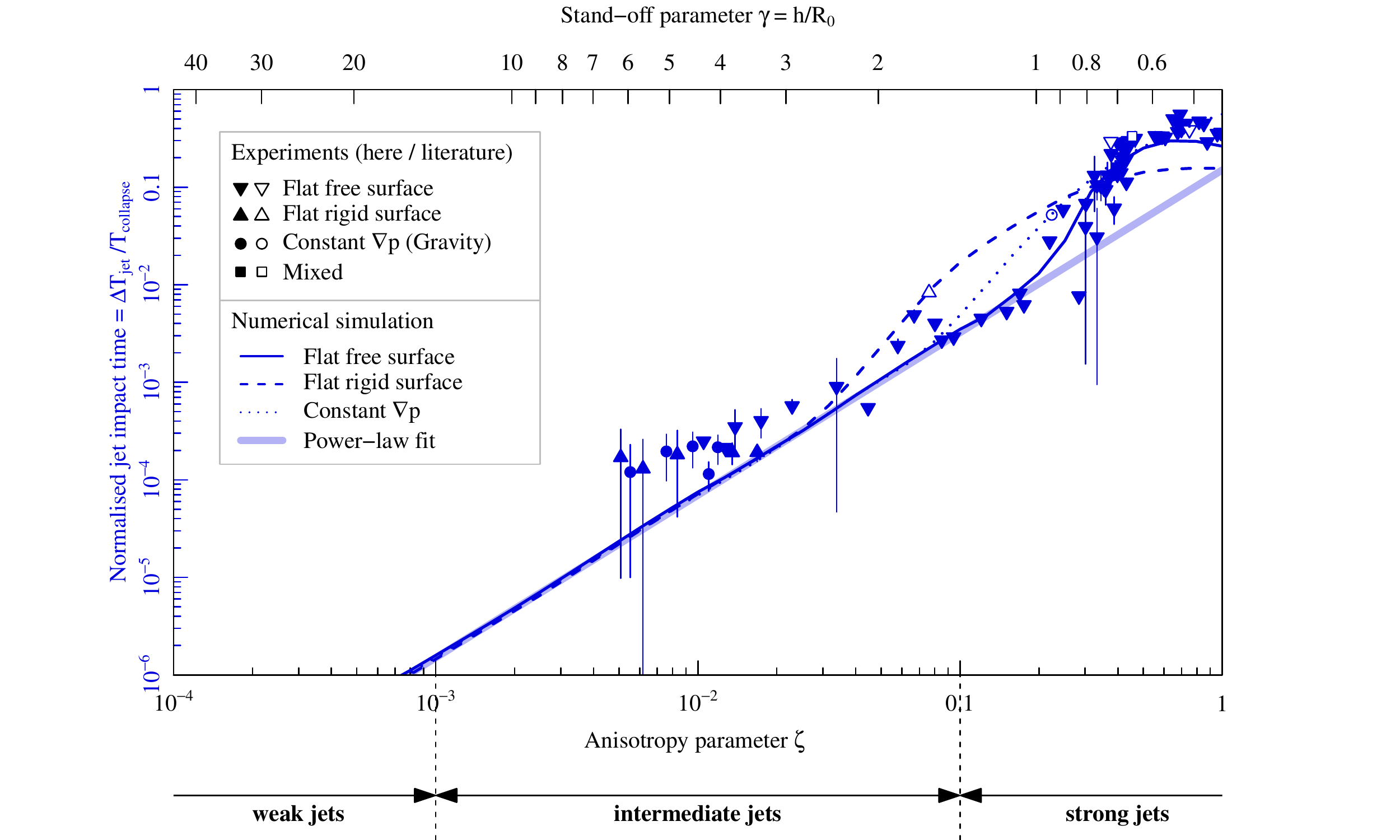}
\caption{ Normalised jet impact time as a function of the anisotropy parameter $\zeta$ and the stand-off parameter $\gamma$. Our experimental data (filled) are compared with literature data (empty): Spark-induced bubbles subject to buoyancy, $R_{0} \sim 50$~mm \citep{Zhang2015}, spark-induced bubbles near a free surface and a rigid surface, $R_{0} \sim 10$~mm~\citep{Zhang2013}, lens-based laser-induced bubbles near a rigid surface, $R_{0} = 1.45$~mm \citep{Philipp1998}. The dotted, dashed and solid lines are the numerical models with a constant pressure gradient, near a rigid surface and near a free surface respectively. The thick line is the power-law fit in equation~(\ref{eq:powerless}), discussed in section \ref{s:interp}.}
\label{fig:jettime}
\end{center}
\end{figure}

An interesting parameter characterising a micro-jet is the moment at which the jet pierces the opposite bubble wall during the collapse.
The normalised jet impact time is defined as $\Delta T_{\rm jet}/T_{\rm collapse}$, where $\Delta T_{\rm jet}$ is the time interval from the jet impact to the collapse point (i.e.\ the minimal radius of the toroidal bubble), and $T_{\rm collapse}$ is the time interval from the maximal bubble volume to the collapse point. 
The timing of the jet impact is measured through high-speed visualisations either by observing the moment at which a shock wave is emitted due to the impact, such as in figures~\ref{fig:strongshock} and~\ref{fig:intermediateshock}, or by looking at the bubble interior for the more obvious cases.

Figure~\ref{fig:jettime} displays the normalised jet impact time as a function of $\zeta$ and $\gamma$. 
It is evident that the jet pierces the bubble at an earlier stage in the collapse with increasing $\zeta$, i.e.\ as the bubble deformation becomes more pronounced.
In the most deformed cases the jet can pierce the bubble as early as at half of the collapse time. 
On a linear scale, this parameter varies predominantly in the strong jet regime, but all jets that pierce the bubble (i.e.\ in strong and intermediate regimes) do so before the collapse. 
In the intermediate regime, however, the jet impact occurs very close to the collapse moment, i.e.\ $\Delta T_{\rm jet}/T_{\rm collapse}<1\%$. 
This is, in fact, how we chose the dividing value $\zeta = 0.1$ between intermediate and strong jets. 
The offset between data and model around $\zeta=0.01$ is probably attributed to difficulties of measuring normalised jet impact times below $10^{-4}$, skewing the existing data points towards higher values.

In the simulation, we calculate the evolution of the surface of the simply connected bubble up to the moment of jet impact using the boundary integral method explained in section~\ref{s:num}. Beyond this instant, the collapse time of the torus is calculated using the vortex ring model~\citep{Wang2005}, where the complex shape of the vortex ring is approximated by a circular torus of identical volume, mean radius, circulation $\Gamma$ and initial collapse speed.
The collapse of this torus is computed using equation~(8) in~\cite{Chahine1983}
\footnote{Note that the torus collapse time given in eq.~(12) of this reference is not sufficient for this purpose, since it neglects the significant initial collapse speed and circularity of the torus.}.
The numerical calculations agree with the experimental results within their uncertainties.
These models are almost identical for the different jet drivers up to about $\zeta = 0.03$, with major differences arising in the strong jet regime, in particular for the rigid surface.
These discrepancies are probably attributed to the more pronounced differences in the bubble geometries between the different jet drivers, as seen in figure~\ref{fig:zoom} for example at $\zeta = 0.1$.
As a consequence, whether the jet impacts on a single point (rigid surface) or on an annular ring (free surface) lead to a different volume of the remaining toroidal bubble, which in turn leads to a longer collapse time.


\subsection{Jet speed} 
\label{s:jetspeed}

\begin{figure}
\begin{center}
\includegraphics[width=\textwidth, trim=1.9cm 0cm 3cm -0.3cm, clip]{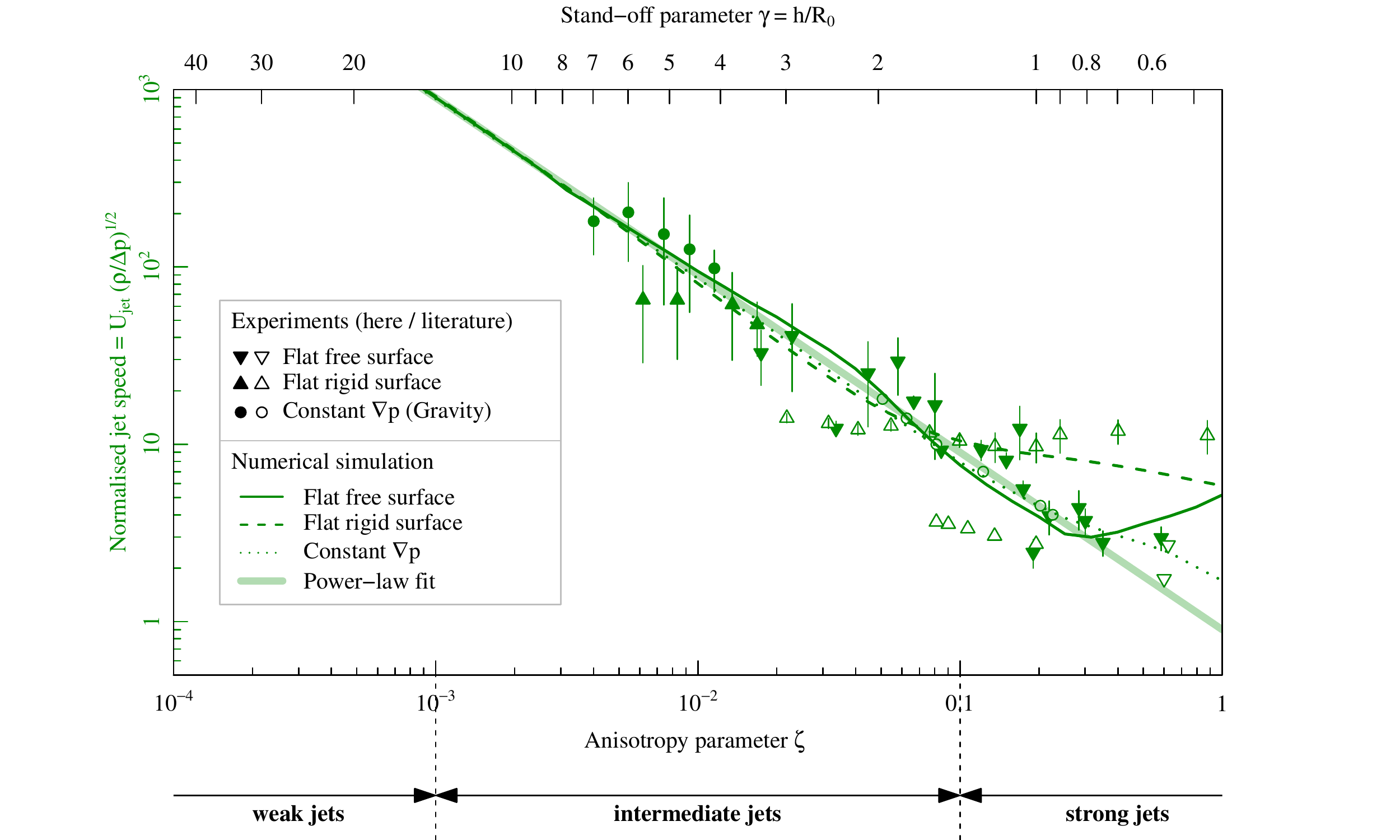}
\caption{ Normalised jet speed as a function of the anisotropy parameter $\zeta$ and the stand-off parameter $\gamma$. Our experimental data (filled) are compared with literature data (empty): Spark-induced bubbles subject to buoyancy, $R_{0} \sim 50$~mm \citep{Zhang2015}, spark-induced bubbles near a free surface and a rigid surface, $R_{0} \sim 10$~mm \citep{Zhang2013}, lens-based laser-induced bubbles near a rigid surface, $R_{0} = 1.45$~mm \citep{Philipp1998}, lens-based laser-induced bubbles near a free surface, $R_{0} \sim 1.3$~mm \citep{Robinson2001}, lens-based laser-induced bubbles near a rigid surface, $R_{0} = 1.55$~mm \citep{Brujan2002}, focused ultrasound-induced bubbles near a rigid surface, $R_{0} = 200$~$\mu$m \citep{Brujan2005}. The dotted, dashed and solid lines are the numerical models with a constant pressure gradient, near a rigid surface and near a free surface respectively. The thick line is the power-law fit in equation~(\ref{eq:powerless}), discussed in section \ref{s:interp}.}
\label{fig:jetspeed}
\end{center}
\end{figure}

An important parameter that describes the micro-jet dynamics is the jet speed. 
Here we define it as the maximum jet speed before the impact on the opposite bubble wall, normalised by the characteristic speed $(\Delta p/\rho)^{1/2}$~\citep{Plesset1970}. 
The speed is measured from visualisations of the bubble interior, where the jet is visible inside the bubble prior to the impact (such as in figure~\ref{fig:strong}).

Figure~\ref{fig:jetspeed} displays our measurements of the normalised jet speed as a function of $\zeta$ and $\gamma$, together with selected data from the literature.
They reveal a decrease of the normalised jet speed with increasing $\zeta$.
This is explained by the jet piercing the bubble earlier at high $\zeta$ (as seen in section~\ref{s:jetimpact}), when the bubble interface speed is still relatively low. 
In fact, the jet speed tends to infinity as $\zeta \rightarrow 0$, i.e.\ as we approach the limit of spherical collapse in the Rayleigh theory.
It should be noted that we are unable to measure jet velocities for $\zeta < 3\cdot 10^{-3}$ with our temporal and spatial resolution.

The measurements for gravity- and free surface-driven jets are in good agreement with the numerical simulations. However, the data points drawn from the literature~\citep{Philipp1998,Brujan2002} for jets induced by a rigid surface appear to deviate from the corresponding model at $\gamma>2$ and $\gamma<1$. The reasons for this deviation are not entirely clear, but we note that the value of the jet speed depends sensibly on when exactly the measurement is performed. Besides, extracting jet speeds from high-speed images is a challenge, as it requires a highly transparent bubble interface to see the bubble interior in addition to sufficient spatial and temporal resolutions. Another potential caveat with these observations is the optical refraction on the bubble surface. It should be noted that in reality jets are expected to stop accelerating once they approach the speed of sound of the liquid and the potential flow theory starts to fail. This is typically at $\zeta<0.01$ in standard water conditions (where $(\Delta p/\rho)^{1/2}\approx10\,{\rm m\,s^{-1}}$, hence $U_{\rm jet}\gtrsim900\,{\rm m\,s^{-1}}$).

Interestingly, in the weak jet regime (where we only have model data) and in the intermediate jet regime up to $\zeta=0.1$, the jet speed is entirely set by $\zeta$ with negligible dependence on the jet driver. Only for asymmetries larger than $\zeta=0.1$ can we notice a significant deviation of jets associated with a rigid surface relative to those associated with a free surface and/or gravity.

\subsection{Bubble displacement} 
\label{s:deltay}

\begin{figure}
\begin{center}
\includegraphics[width=\textwidth, trim=1.9cm 0cm 3cm -0.3cm, clip]{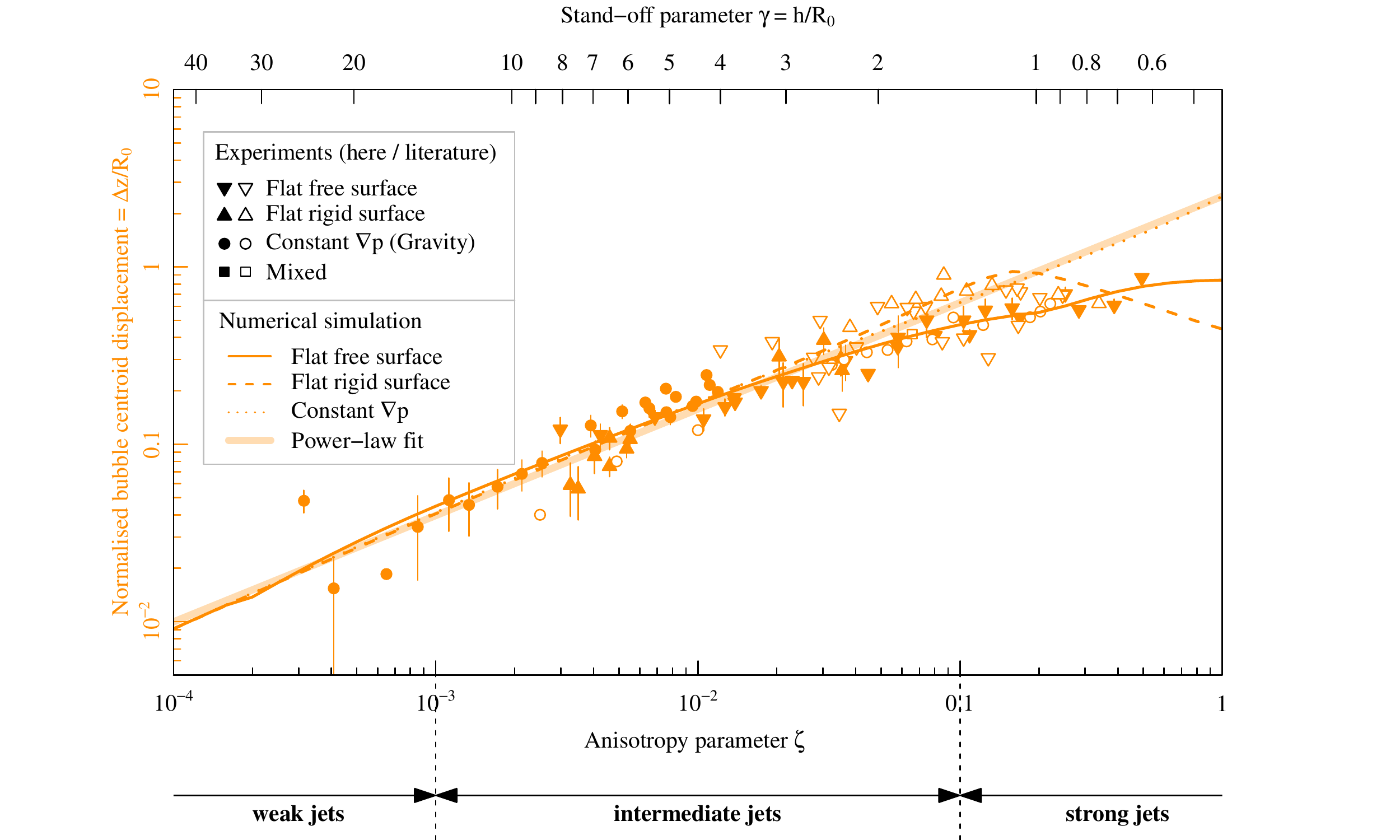}
\caption{ Normalised bubble centroid displacement from generation to collapse as a function of the anisotropy parameter $\zeta$ and the stand-off parameter $\gamma$. Our experimental data (filled) are compared with literature data (empty): Spark-induced bubbles subject to buoyancy, $R_{0} \sim 50$~mm \citep{Zhang2015}, spark-induced bubbles near a free surface and a rigid surface, $R_{0} \sim 10$~mm \citep{Zhang2013}, lens-based laser-induced bubbles near a rigid surface, $R_{0} = 1.55$~mm \citep{Brujan2001b}, underwater explosion bubble subject to buoyancy $R_{0} = 0.54$~m \citep{Hung2010}, underwater explosion bubble near a free surface $R_{0} \sim 0.17$~m \citep{Klaseboer2005}, lens-based laser-induced bubbles near a rigid surface, $R_{0} = 0.65$~mm \citep{Tomita2002,Tomita2003}, lens-based laser-induced bubbles near a rigid and a free surface, $R_{0} \sim 1.5$~mm \citep{Gregorcic2007}. The dotted, dashed and solid lines are the numerical models with a constant pressure gradient, near a rigid surface and near a free surface respectively. The thick line is the power-law fit in equation~(\ref{eq:powerless}), discussed in section \ref{s:interp}.}
\label{fig:centroid}
\end{center}
\end{figure}

Another jet parameter worth discussing is the bubble centroid displacement. 
While not strictly a micro-jet property, this displacement is the most straightforward way to detect a Kelvin impulse.
The bubble displacement $\Delta z$ is defined as the distance travelled by the bubble centroid between the bubble generation and collapse, in the rest-frame of the liquid. 
 Special care is required when the bubble splits into multiple parts at higher pressure anisotropies. Here we define the centroid position at the collapse as the position of the jet tip at its impact onto the opposite bubble wall.
The experimental results for centroid displacement presented here are normalised by the bubble maximum radius, $\Delta z/R_{0}$. 
Note that some authors choose to normalise $\Delta z$ by the distance $h$ from the flat surface, but this normalisation would not be applicable to other causes of micro-jets such as gravity.

Our measurements of $\Delta z/R_{0}$ are shown in figure~\ref{fig:centroid} as a function of $\zeta$ and $\gamma$, together with selected data from literature. 
In general, we find good agreement between the data points from the different jet drivers, within the measurement uncertainties. 
Overall, we find an increase of the normalised centroid motion with increasing $\zeta$. 
A particularly important finding is that even in the weak jet regime, where the jet speed, impact time and volume (as we will see in section~\ref{s:jetsize}) become cumbersome parameters to measure experimentally, the displacement remains a significant and measurable quantity as evidenced in figure~\ref{fig:centroid}. The larger scatter of the literature data (empty symbols) might be attributed to the fact that the definition of ``collapse position'' or ``centre of minimum bubble volume'' is not always clear for a strongly deformed bubble and therefore the data extraction may not have been done in the same way in all experiments.

The numerical models agree well with the empirical data. In the weak and intermediate jet regimes up to about $\zeta=0.1$, the simulated displacement shows little dependence on the jet driver and is thus almost entirely dictated by the value of $\zeta$. 
For asymmetries larger than $\zeta=0.1$, the displacement starts to depend significantly on whether the anisotropy is associated with a rigid surface, free surface or gravity.

\subsection{Bubble volume at jet impact} 
\label{s:vtor}

\begin{figure}
\begin{center}
\includegraphics[width=\textwidth, trim=1.9cm 0cm 3cm -0.3cm, clip]{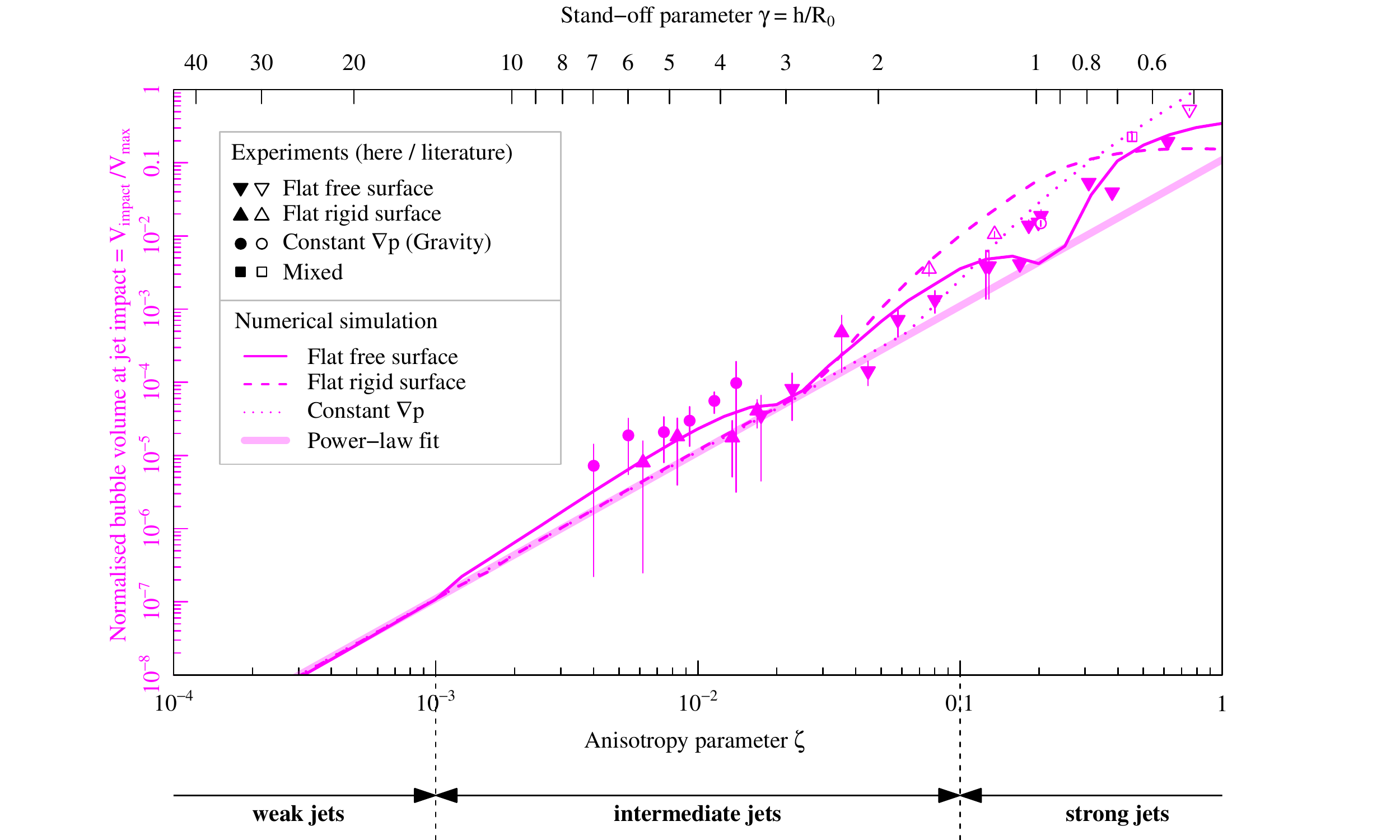}
\caption{ Normalised bubble volume at jet impact as a function of the anisotropy parameter $\zeta$ and the stand-off parameter $\gamma$. Our experimental data (filled) are compared with literature data (empty): Spark-induced bubbles subject to buoyancy, $R_{0} \sim 45$~mm \citep{Zhang2015}, spark-induced bubbles near a free surface and a rigid surface, $R_{0} \sim 10$~mm \citep{Zhang2013}, lens-based laser-induced bubbles near a rigid surface, $R_{0} = 1.45$~mm \citep{Philipp1998}. The dotted, dashed and solid lines are the numerical models with a constant pressure gradient, near a rigid surface and near a free surface respectively. The thick solid line is the power-law fit in equation~(\ref{eq:powerless}), discussed in section \ref{s:interp}.}
\label{fig:vtor}
\end{center}
\end{figure}

The bubble volume $V_{\rm impact}$ at the jet impact is yet another interesting parameter characterising the jet formation. It is a more easily definable size-parameter than the jet-size itself. The normalised bubble volume at jet impact is defined as $V_{\rm impact}/V_{\rm max}$, where $V_{\rm max} = (4\pi/3)R_{0}^3$.
Experimentally, $V_{\rm impact}=2\pi xA$ is obtained from the high-speed visualisations by measuring the area $A$ of the toroid-cross section (averaged between the two cross-sections seen on either side of the jet-axis) and the distance $x$ between the geometric central line of the toroid and the jet-axis. 

Figure~\ref{fig:vtor} shows the normalised bubble volume at jet impact as a function of $\zeta$ and $\gamma$.
This parameter increases with $\zeta$, which is explained by the jet piercing the bubble at an earlier stage during the collapse at higher $\zeta$, when the bubble is still large relative to its final collapse size.
The jets from different drivers follow a similar trend.

The numerical calculations agree well with the empirical data within the uncertainties.
 The different jet drivers exhibit similar trends in the weak and intermediate jet regimes.
The differences, especially in the high-intermediate and strong jet regimes, are explained by the different jet shapes (figure~\ref{fig:simu}, at $\zeta = 0.1-0.3$).
In particular, bubbles collapsing near a free surface produce broad jets that hit the opposite bubble wall on a ring rather than a single point.
In this case, the jet separates the bubble into a smaller bubble \emph{and} a torus, resulting in a more complex bubble shape than a simple torus, which therefore yields a different volume.
This explains the ondulations of the free surface model in figure~\ref{fig:vtor} and makes the bubble volume at jet impact, together with the jet impact timing, the most sensitive parameter to jet drivers.

\subsection{Vapour-jet volume} 
\label{s:jetsize}

\begin{figure}
\begin{center}
\includegraphics[width=\textwidth, trim=1.9cm 0cm 3cm -0.3cm, clip]{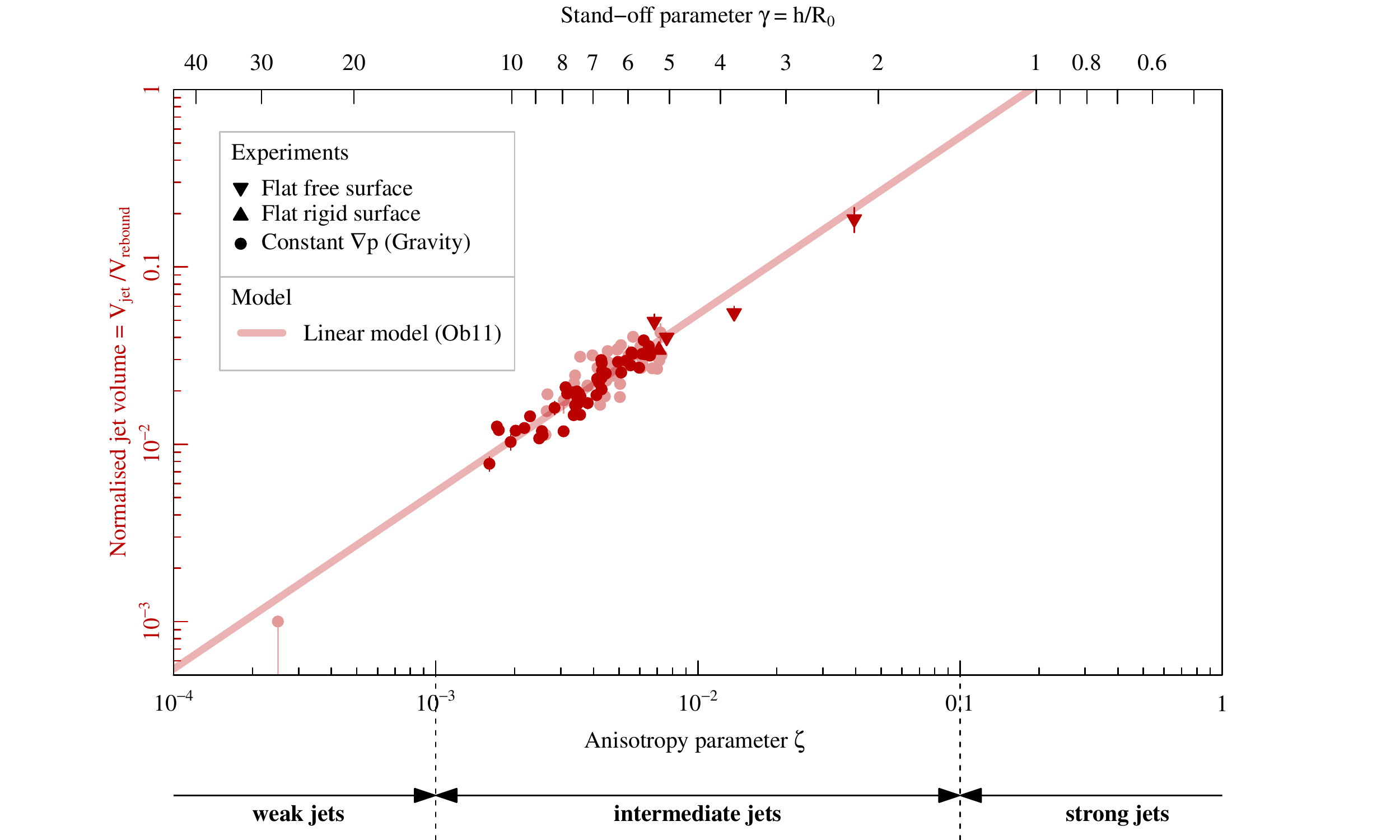}
\caption{ Scaling law for the post-collapse bubble vapour-jet volume. Light data points indicate results from variable gravity ($0g$, $1.2g$, $1.4g$, $1.6g$ and $1.8g$ where $g$=9.81 ms$^{-2}$) and dark points are from normal gravity ($1g$). Maximal bubble radius $R_{0}$ is varied in the range 1-7 mm, liquid pressure $p_{0}$ in the range 8-80 kPa and the dynamic viscosity $\eta$ in the range 1-30 mPa s. The majority of the data points for the constant pressure gradient, and the theoretical model (solid line) in equation~(\ref{eq:jetsize}) are from~\cite{Obreschkow2011}.}
\label{fig:jetscale}
\end{center}
\end{figure}

The final jet parameter discussed in this paper is the post-collapse vapour-jet volume.
The scaling of the vapour-jet volume $V_{\rm jet}$ (figure~\ref{fig:regimes}b), normalised by the rebound volume $V_{\rm rebound}$, as a function of $\zeta$ has been investigated in the intermediate jet regime in~\cite{Obreschkow2011}. 
The data points from this reference are re-plotted in figure~\ref{fig:jetscale}, along with new data for the free surface, as a function of $\zeta$ and $\gamma$. 
The empirical result was a linear relation (thick line in the figure),
\begin{equation}
	\frac{V_{\rm jet}}{V_{\rm rebound}}\approx5.4\zeta,
\label{eq:jetsize}
\end{equation}
valid across a large range of bubble sizes, liquid pressures and viscosities (varied by a factor 30 using glycerol additions). The authors justified the proportionality between $V_{\rm jet}/V_{\rm rebound}$ and $\zeta$ based on Kelvin impulse considerations. They also presented a critical value $\zeta_{c} \approx 4\cdot 10^{-4}$, such that in situations with $\zeta<\zeta_{c}$ the micro-jet does not pierce the bubble wall and no vapour-jet emerges from the rebound bubble. This value is approximately consistent with our choice of $\zeta=10^{-3}$ as the dividing value between the intermediate and weak jet regimes (section~\ref{s:weak}). For a more detailed discussion of the vapour-jet volume, we refer to the original work~\citep{Obreschkow2011}.


\section{Discussion} 
\label{s:discussion}

\subsection{Power-law approximations} 
\label{s:interp}

The dimensionless jet parameters discussed in sections \ref{s:jetimpact}--\ref{s:jetsize} mainly vary with the anisotropy parameter $\zeta$. We also identified a secondary dependence on the jet driver (gravity versus surfaces). According to figures~\ref{fig:jettime}--\ref{fig:jetscale}, this secondary dependence becomes generally negligible in the weak and intermediate jet regimes ($\zeta<0.1$). Furthermore, in these regimes the unique relations between $\zeta$ and the jet parameters appear to be closely matched by power-laws, in particular for the jet speed, the bubble displacement and the vapour-jet volume. A chi-square fit to the simulated models over the range $\zeta=10^{-4}-0.1$ with uniform weight in $\log(\zeta)$ yields:
\begin{equation}\label{eq:powerless}
\begin{array}{lcll}
\Delta T_{\rm jet}/T_{\rm collapse} &=& 0.15\,\zeta^{5/3} & \text{(normalised jet impact time)}\\
U_{\rm jet}/(\Delta p/\rho)^{1/2} &=& 0.9\,\zeta^{-1} & \text{(normalised jet speed)}\\
\Delta z/R_{0} &=& 2.5\,\zeta^{3/5} & \text{(normalised bubble displacement)}\\
V_{\rm impact}/V_{\rm max} &=& 0.11\,\zeta^{2} & \text{(normalised bubble volume at jet impact)}\\
V_{\rm jet}/V_{\rm rebound} &=& 5.4\,\zeta & \text{(normalised volume of vapour-jet)}
\end{array}
\end{equation}

\begin{figure}
\begin{center}
\includegraphics[width=\textwidth, trim=0.3cm 0cm 0.45cm -0.3cm, clip]{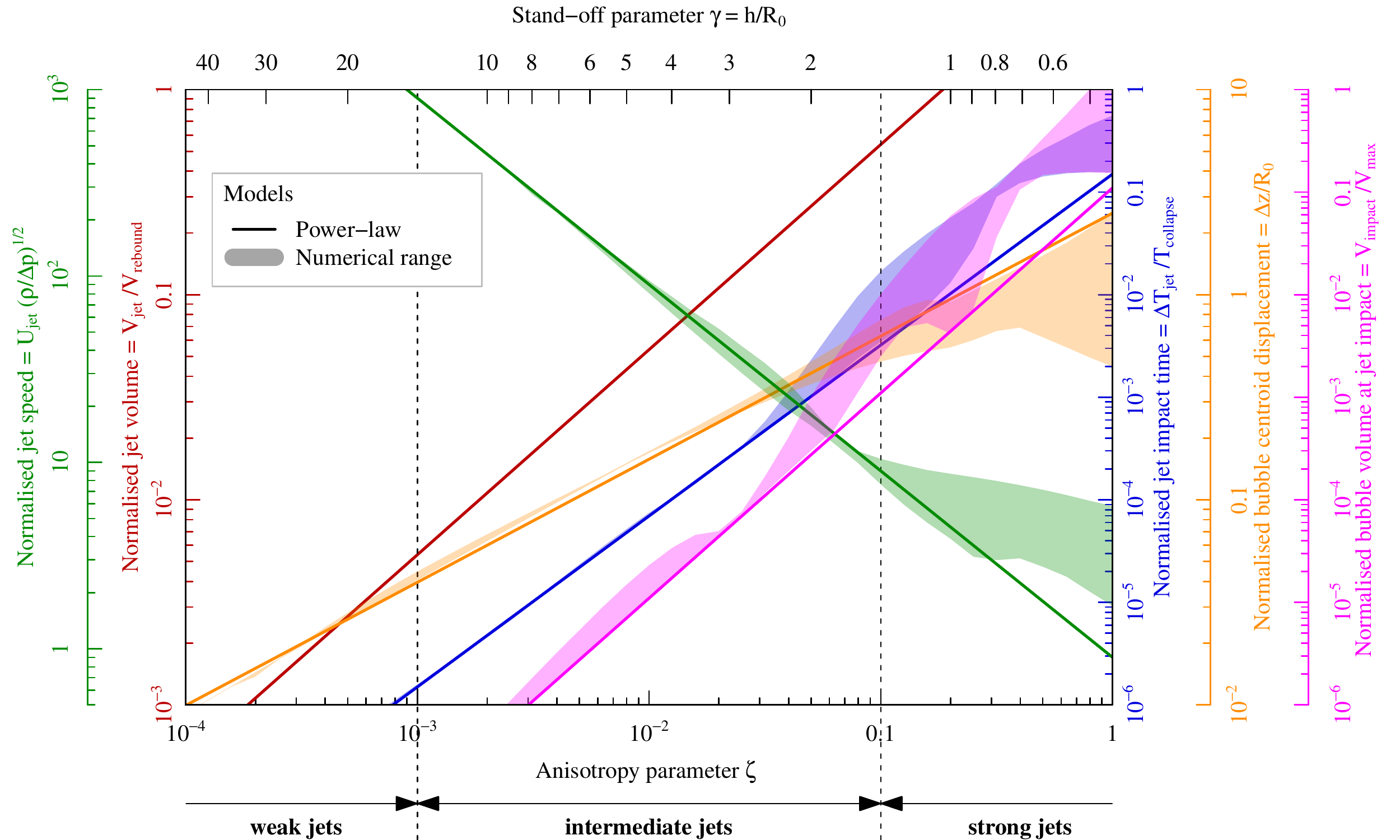}
\caption{ Summary of the micro-jet parameters across all regimes. The power-laws for the normalised jet impact time, jet speed, bubble centroid displacement, bubble volume at jet impact and vapour-jet volume~\citep{Obreschkow2011} are plotted as a function of the anisotropy parameter $\zeta$ and the stand-off parameter $\gamma$. The shaded areas describe the range spanned by the different jet drivers, which is calculated numerically (see figures~\ref{fig:jettime}-\ref{fig:vtor}).}
\label{fig:main}
\end{center}
\end{figure}

The last relation is not a fit to numerical models, but the empirical equation (\ref{eq:jetsize}), repeated for completeness. These power-laws are represented by the thickest lines in figures~\ref{fig:jettime}--\ref{fig:jetscale} and are synthesised in figure~\ref{fig:main} together with the range of numerical results spanned by various jet drivers (shaded regions). The power-laws provide a simple tool to predict the dynamics of an aspherical bubble collapse in a large range of conditions, without the need for complex computations.

To understand the reasons for this power-law behaviour and explain the power-law exponents, we recall that power-laws are generally an expression of scale-free behaviour. 
Scale-free means that the physical system is geometrically similar, independently of its overall scale. 
Of course, the \textit{whole evolution} of a jetting bubble is \textit{not} scale-free across a range of $\zeta$, because the maximum bubble radius is independent of $\zeta$, while the jet parameters vary with $\zeta$. Approximate scale-freeness can, however, be found at the \textit{single instant} when the jet impacts on the opposite side of the bubble wall (blue lines in figure~\ref{fig:simu}). 
For small values of $\zeta$ ($\zeta<0.1$), the bubble at this instant has a universal, bowl-like shape. 
Only the \textit{size} varies with $\zeta$, but the bubble \textit{shape} is independent of the value and the origin of $\zeta$.

Scale-freeness at the jet impact stage means that all lengths scale proportionally to the characteristic bubble radius $r\equiv R(t)/R_{\rm 0}$ at this stage. 
Corresponding volumes and masses scale as $r^{3}$. 
To find the characteristic scaling of velocities, we note that, for small $\zeta$, the bubble deformation occurs very late in the collapse phase (i.e. $r\ll1$). 
In this phase, the time-evolution of the bubble radius satisfies $\dot{r}=r^{-3/2}$, which is the asymptotic behaviour of the Rayleigh equation as $r\rightarrow 0$~\citep{Obreschkow2012}. 
Given that masses scale as $r^{3}$ and velocities as $r^{-3/2}$, linear momentum (= product of mass and velocity) scales as $r^{3}\dot{r}=r^{3/2}=\dot{r}^{-1}$. 
Since the momentum of the bubble is proportional to $\zeta$ (see equation~(\ref{eq:Igradient})), we find $r\sim \zeta^{2/3}$ (see figure~\ref{fig:zoom}) and $\dot{r}\sim \zeta^{-1}$. 
This explains the numerical scalings $V_{\rm impact}\sim \zeta^{2}$ and $U_{\rm jet}\sim \zeta^{-1}$.

The asymptotic equation of the spherical collapse $\dot{r}=r^{-3/2}$ solves to $r\sim \tilde{t}^{2/5}$, where $\tilde{t}=1-t$ is the time backwards from the collapse point, normalised to the collapse time~\citep{Obreschkow2012}. 
Thus, for small $\zeta$, we expect $\Delta T_{\rm jet}\sim r^{5/2}\sim (\zeta^{2/3})^{5/2}=\zeta^{5/3}$, as confirmed by the numerical simulation.

Our interpretation of the vapour-jet scaling is more speculative, since we did not simulate the formation of this jet. 
One might naively expect the volume of the vapour-jet $V_{\rm jet}$ to scale as $r^{3}\sim \zeta^{2}$, just like $V_{\rm impact}$. 
However, the vapour-jet is not a feature at the instant of the jet impact. 
Hence the arguments of scale-freeness of the previous paragraphs do not apply. 
The correct reasoning is that the volume of the vapour-jet is the part of the micro-jet that actually gets pushed through the bubble wall during the time interval of the rebound. 
The vapour-jet volume therefore depends both on the characteristic micro-jet volume and on the jet speed. 
Consequently, we expect $V_{\rm jet}\sim r^{3}\dot{r}\sim \zeta^{2}\zeta^{-1}=\zeta$, in agreement with the experimental results.
This explanation should be tested against more detailed modelling of the vapour-jet formation in future work.

Finally, the normalised displacement of the bubble centroid $\Delta z$ is expected to scale as $\Delta z\sim r\sim \zeta^{2/3}$, if this displacement occurs 
uniquely at the final collapse stage, where the scale-free picture applies. 
The power-law exponent of $2/3=0.666\dots$ is indeed the best fit to the simulations for very small values of $\zeta$ ($\zeta<10^{-3}$), where almost all the bubble motion occurs just before and after the final collapse 
point. 
However, for larger values of $\zeta$, a non-negligible fraction of the bubble motion occurs at larger bubble radii, where $|\dot{r}|<r^{-3/2}$ according to eq. (7) in~\cite{Obreschkow2012}. 
Hence, the power-law index between $\Delta z$ and $\zeta$ must drop below $0.666…$. 
This prediction is consistent with our numerical finding that $\Delta z$ scales approximately as $\Delta z \sim \zeta^{0.6}=\zeta^{3/5}$ over the range $\zeta<0.1$.

\subsection{Application of scaling relations} 

\begin{figure}
\begin{center}
\includegraphics[width=\textwidth, trim=0.3cm 0cm 0.45cm -0.3cm, clip]{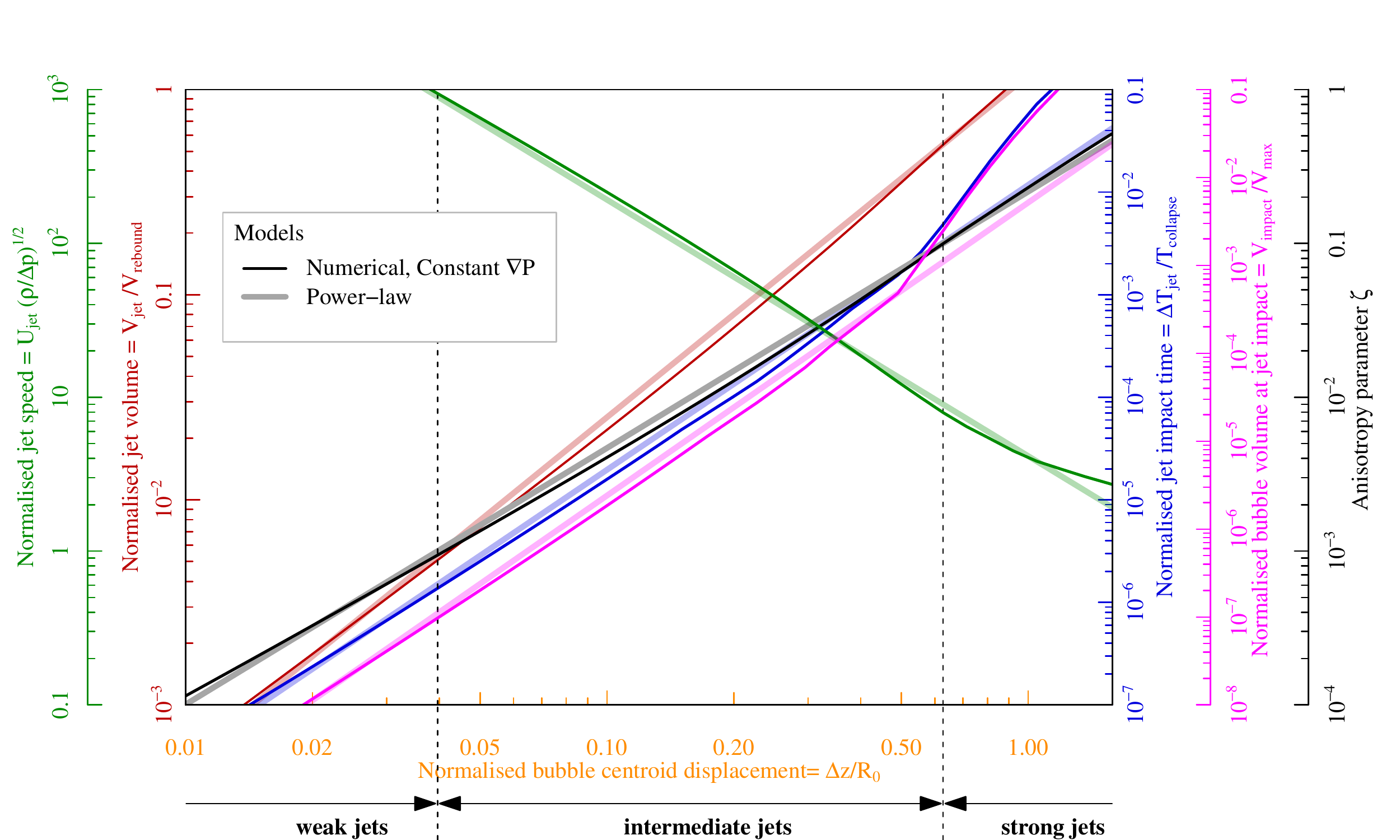}
\caption{The pressure anisotropy parameter $\zeta$ and the normalised jet impact time, the normalised jet speed, the normalised bubble volume at jet impact and the normalised jet volume~\citep{Obreschkow2011} are plotted as a function of the normalised bubble centroid displacement for jets driven by a constant pressure gradient. The simulated models and the power-law fits are plotted with dark and light lines, respectively.}
\label{fig:deltay}
\end{center}
\end{figure}

 The power-laws are a useful predictive tool of the micro-jet physics in known pressure anisotropies $\zeta < 0.1$.
In the strong jet regime ($\zeta > 0.1$) (and in the high-intermediate regime for the jet impact time and bubble volume at jet impact), more accurate, non-linear scaling relations can be obtained numerically for specific jet drivers, as shown in figures~\ref{fig:jettime}--\ref{fig:jetscale} and tabulated in Appendix~\ref{app:data}.

An interesting consequence of the jet-scalings with $\zeta$ is that one may reciprocally use a known jet observable to estimate the pressure anisotropy in which the bubble is collapsing.
Consequently, the measurement of a single jet observable suffices to estimate the rest of the parameters.
The bubble centroid displacement, for instance, presents the advantage of being the easiest measurable quantity of an aspherical bubble collapse across a large range of pressure anisotropies.
It therefore serves as a simple and useful predictor of the full micro-jet physics.
As an example, the particular case of jets driven by a constant pressure gradient $\gradp$ is presented in figure~\ref{fig:deltay}, where the various jet parameters and the anisotropy parameter $\zeta$ are plotted as a function of the bubble displacement $\Delta z/R_{0}$. For reference, we also show the results corresponding to the simple power-laws. 
Their similarity in the weak and intermediate regime ($\zeta < 0.1$) implies that figure~\ref{fig:deltay} would look nearly the same for other jet drivers in this regime.

\subsection{Limitations} 

Let us conclude this discussion by addressing a few limitations of the unified perspective offered by the single anisotropy parameter $\zeta$. 
As mentioned before, the micro-jets in the strong jet regime, where more complex jet morphologies are produced, cannot be fully described by $\zeta$ independently of the jet drivers. 
At these high anisotropies, strong variations in the jet parameters for different jet origins occur as a direct consequence of the higher-order terms in equation~(\ref{eq:p}), as discussed in section~\ref{s:strong}.
Predictions in this regime should be made numerically for the specific jet drivers.

Combining the effect of multiple jet drivers generally produces jets that follow the same scaling laws as jets from a single driver in the weak and intermediate jet regimes. 
However, attention should be paid to situations where several strong jet drivers act simultaneously in opposite directions \citep[e.g.\ gravity and rigid boundary in][]{Zhang2015}, as they may yield a low resultant $\zeta$ although the higher-order terms in equation~(\ref{eq:p}) remain significant. This can result in bubble splitting, producing e.g.\ the ``hourglass'' bubble~\citep{Blake1987}, the dynamics of which cannot be predicted by our approach.

So far, our investigations have mainly focused on flat rigid or free surfaces. Curved \citep{Tomita2002}, flexible \citep{Brujan2001} and composite \citep{Tomita2003} surfaces would require specific corrections to $\boldsymbol{\zeta}$ in eq.~(\ref{eq:zeta}), which would serve as an interesting addition to the diverse family of micro-jets. 
Furthermore, as a consequence of the assumption that viscosity and surface tension play a minor role in the micro-jet dynamics, our approach is limited to bubbles of a certain scale in water and we do not account for jets produced by capillary phenomena. 
Viscosity and surface tension that become important in e.g.\ biomedical applications that deal with micron-sized bubbles in viscous liquids, break the scale-freeness and may change the trends with $\zeta$. 
It would be an interesting opening for future work.

Finally, it should be noted that the lifetime of bubbles investigated in the present study includes the bubble growth, which strongly affects the subsequent motion (in particular for bubbles near a flat surface at $\gamma<1$). Our numerical tool (see section~\ref{s:num}) provides the option to exclude the growth phase and start with a perfectly spherical bubble at its maximal radius.

\section{Conclusion} 

In this work, we conducted a qualitative and quantitative analysis of the micro-jet dynamics of a single cavitation bubble in a large range of conditions. 
By introducing a dimensionless anisotropy parameter $\zeta$, we arrived at a unified framework describing micro-jets of virtually any strength, caused by various jet drivers, in particular gravity, free surfaces, rigid surfaces and combinations thereof. This successful unification of the micro-jet family through $\zeta$, a normalised version of the Kelvin impulse, fosters Blake's~\citep{Blake2015} view that the Kelvin impulse is a ``fundamental [...] enormously valuable concept''.

The main contribution of this work is the realisation that, in normalised coordinates, $\zeta$ fully defines the jet-physics, once the jet driver (e.g.\ gravity or nearby boundaries) has been identified. Furthermore, for small Kelvin impulses ($|\mathbf{I}|<R_0^3 \sqrt{\Delta p\rho}/2$, that is for $\zeta<0.1$) the jet physics becomes virtually independent of the jet driver. This powerful aspect of the Kelvin impulse comes about despite -- or rather because of -- the concerns raised by \cite{Lauterborn1982} about this impulse being an integral value.

We have investigated, both experimentally and numerically, how different jet characteristics vary with $\zeta$.
The normalised jet impact time, the jet speed, the bubble centroid displacement, the bubble volume at jet impact and the vapour-jet volume can all be approximated by power-laws of $\zeta$ up to $\zeta \approx 0.1$, independently of the jet drivers. 
A single observable may be used to predict another jet parameter or estimate the pressure anisotropy, as shown in figure~\ref{fig:deltay}.

The micro-jets have been phenomenologically classified into three distinct regimes: weak, intermediate and strong jets. 
We showed that such a categorisation presents a useful thinking tool to distinguish visually very different jets, which nonetheless all fit in the unified framework of the $\zeta$ parameter.
Weak jets ($\zeta < 10^{-3}$) hardly pierce the bubble, but remain within the bubble throughout the collapse and rebound. 
Intermediate jets ($10^{-3} < \zeta < 0.1$) pierce the opposite bubble wall very late in the collapse phase and clearly emerge during the rebound. 
Strong jets ($\zeta > 0.1$) pierce the bubble significantly before the moment of collapse and their dynamics is strongly dependent on the jet driver.

The presented results might serve as a step towards unifying the quickly diversifying research field of cavitation and towards reaching a unified framework for the energy distribution between all collapse-related phenomena.
A precise control of the power of micro-jets would allow, for instance, the attenuation of detrimental jet-induced erosion as well as the targeting of cancerous cells or highly localised drug delivery. Such new research avenues may benefit from the framework and predictive tools presented here.\\

\noindent\textbf{Acknowledgements} We gratefully acknowledge the support of the Swiss National Science Foundation (Grant no.~513234), the support of the University of Western Australia Research Collaboration Award obtained by co-authors DO and MF, and the support of the European Space Agency.

\bibliographystyle{jfm}
\bibliography{microjets}

\newpage
\appendix

\section{Mathematical derivations}\label{app:derivations} 
The evolution of a spherical bubble of radius $R$ in a liquid of density $\rho$ and constant over-pressure $\Delta p$ (relative to the bubble content) is governed by the Rayleigh equation~\citep{Rayleigh1917}
\begin{equation}\label{eq_rayleigh}
	\frac{3}{2}\left(\frac{{\rm d}R}{{\rm d}T}\right)^2+\frac{{\rm d}^2 R}{{\rm d} T^2}R = -\frac{\Delta p}{\rho}.
\end{equation}
We can define the time $T$ such that the bubble is at the maximal radius $R_0$ at $T=0$. Equation~(\ref{eq:rayleigh}) then implies that the radius vanishes at $T=\pm T_{\rm c}$, where $T_{\rm c}=\xi R_0(\rho/\Delta p)^{1/2}$ and $\xi$ is a numerical constant, called the Rayleigh factor. Upon normalising the radius to $r\equiv R/R_0\in[0,1]$ and the time to $t\equiv T/T_{\rm c}\in[-1,1]$, the Rayleigh equation can be simplified to a dimensionless first order differential equation~\citep{Obreschkow2012},
\begin{equation}\label{eq:rayleigh}
	\left(\frac{{\rm d}r}{{\rm d}t}\right)^2=\frac{2}{3}\xi^2\left(r^{-3}-1\right).
\end{equation}
Taking the square-root on both sides (with minus sign on the RHS), and integrating $t=0...1$ and $r=1...0$, this equation readily solves to
\begin{equation}\label{eq:master2}
	\int_0^1 f\,{\rm d}t = \sqrt{\frac{3}{2}}\xi^{-1}\int_0^1 \frac{f\,{\rm d}r}{\sqrt{r^{-3}-1}},
\end{equation}
for any time-dependent function $f$.
Upon performing the substitution $s\equiv r^3$ (hence ${\rm d}r=\frac{1}{3}s^{-2/3}{\rm d}s$), we get
\begin{equation}\label{eq:master}
	\int_0^1 f\,{\rm d}t = \frac{1}{\sqrt{6}\xi}\int_0^1 \frac{f\,{\rm d}s}{s^{1/6}\sqrt{1-s}}.
\end{equation}
Equation~(\ref{eq:master}) is the central equation, from which we can derive the collapse time and various instances of the Kelvin impulse.

\subsection*{Collapse time} 
To get the Rayleigh factor $\xi$, it suffices to set $f=1$ in Equation~(\ref{eq:master}). The LHS then becomes $\int_0^1{\rm d}t=1$, and hence
\begin{equation}
	\xi = \frac{1}{\sqrt{6}}\int_0^1 \frac{{\rm d}s}{s^{1/6}\sqrt{1-s}} = \frac{1}{\sqrt{6}}B\bigg(\frac{5}{6},\frac{1}{2}\bigg)\approx{0.9146813565},
\end{equation}
where $B(x,y)\equiv\int_0^1 t^{x-1}(1-t)^{y-1}{\rm d}t$ is the beta-function.

\subsection*{Kelvin impulse of a bubble in an external pressure gradient} 
Let us start with Blake's equation~\citep{Blake2015} for the momentum (Kelvin impulse) acquired by the liquid during the growth and collapse of a spherical bubble in a constant pressure gradient,
\begin{equation}\label{eq:Igradpblake}
	\mathbf{I} = \gradp \int_{-T_{\rm c}}^{T_{\rm c}}V {\rm d}T,
\end{equation}
where $V$ is the volume of the bubble at time $T$. (Note that Blake presents this equation for the particular case of a gravity-driven gradient $|\gradp|=\rho g$ and he uses the different convention that the bubble is generated at $T=0$ and collapses at $T_{\rm c}$.) Equation~(\ref{eq:Igradpblake}) can be rewritten as
\begin{equation}
	\mathbf{I} = 2\gradp \int_{0}^{T_{\rm c}}\!\!V {\rm d}T
	= \frac{8\pi}{3}\gradp \int_{0}^{T_{\rm c}}\!\!\!R^3 {\rm d}T
	= \frac{8\pi}{3}T_{\rm c}R_0^3\gradp \int_{0}^{1}r^3 {\rm d}t
	= \frac{8\pi\xi}{3}R_0^3(\Delta p\rho)^{1/2}\boldsymbol{\zeta}\int_{0}^{1}r^3 {\rm d}t.
\end{equation}
To evaluate the integral on the RHS we use equation~(\ref{eq:master}) with $f=r^3\equiv s$,
\begin{equation}
	\int_{0}^{1}r^3 {\rm d}t = \frac{1}{\sqrt{6}\xi}\int_0^1 \frac{s\,{\rm d}s}{s^{1/6}\sqrt{1-s}} = \frac{B(11/6,1/2)}{B(5/6,1/2)} = \frac{5}{8}.
\end{equation}
Hence,
\begin{equation}\label{eq:Igradientexact}
	\mathbf{I} = \frac{5\pi}{3\sqrt{6}}B\bigg(\frac{5}{6},\frac{1}{2}\bigg)R_0^3\sqrt{\Delta p\rho}~\boldsymbol{\zeta}\approx4.789R_0^3\sqrt{\Delta p\rho}~\boldsymbol{\zeta},
\end{equation}
which concludes the derivation of equation~(\ref{eq:Igradient}). Note that $R_0^3\sqrt{\Delta p\rho}$ has the dimension of momentum, as required.

\subsection*{Kelvin impulse of a bubble near a rigid/free surface} 
Blake~\citep{Blake2015} also derives the equation of the Kelvin impulse for a bubble near a rigid or free surface,
\begin{equation}
	|\mathbf{I}_{\rm surface}| = \frac{\rho}{16\pi h^2} \int_{-T_{\rm c}}^{T_{\rm c}}(4\pi R^2 \dot{R})^2 {\rm d}T,
\end{equation}
where $h$ is the distance to the rigid or free surface. This expression can be rewritten as
\begin{equation}
	|\mathbf{I}_{\rm surface}| = \frac{2\pi\rho}{h^2} \int_{0}^{T_{\rm c}}\!\!R^4 \dot{R}^2 {\rm d}T
	= \frac{2\pi\rho}{h^2} T_{\rm c}^{-1}R_0^6 \int_{0}^{1}\!\!r^4 \dot{r}^2 {\rm d}t
	= \frac{2\pi}{\xi}(\Delta p\rho)^{1/2} R_0^3 \gamma^{-2}\int_{0}^{1}\!\!r^4 \dot{r}^2 {\rm d}t.
\end{equation}
To evaluate the integral we use equation~(\ref{eq:master}) with $f=r^4\dot{r}^2=\frac{2}{3}\xi^2 s^{4/3}(s^{-1}-1)=\frac{2}{3}\xi^2 s^{1/3}(1-s)$,
\begin{equation}
	\int_{0}^{1}\!\!r^4 \dot{r}^2 {\rm d}t = \frac{2\xi}{3\sqrt{6}}\int_0^1 s^{1/6}(1-s)^{1/2}\,{\rm d}s = \frac{1}{9}B\bigg(\frac{7}{6},\frac{3}{2}\bigg)B\bigg(\frac{5}{6},\frac{1}{2}\bigg).
\end{equation}
Hence,
\begin{equation}\label{eq:Isurfaceexact}
	|\mathbf{I}_{\rm surface}| = \frac{2\pi\sqrt{2}}{3\sqrt{3}}B\bigg(\frac{7}{6},\frac{3}{2}\bigg)R_0^3\sqrt{\Delta p\rho}~\gamma^{-2}\approx0.934R_0^3\sqrt{\Delta p\rho}~\gamma^{-2}
\end{equation}
which concludes the derivation of equation~(\ref{eq:Isurface}). Equating equations~(\ref{eq:Igradientexact}) and (\ref{eq:Isurfaceexact}) yields
\begin{equation}\label{eq:zeta_gamma_exact}
	\zeta = \frac{4B(7/6,3/2)}{5B(5/6,1/2)} \gamma^{-2} \approx 0.195 \gamma^{-2},
\end{equation}
which is the exact expression of equation~(\ref{eq:zeta_gamma}).

\section{Numerical data}\label{app:data} 
\begin{table}
\vspace{-13pt}
\small
\begin{center}
\begin{tabular}{ p{0.7cm} p{0.8cm} p{0.8cm} p{1cm} p{0.8cm} p{0.8cm} p{1cm} p{0.8cm} p{0.8cm} p{1cm} p{0.8cm} p{0.8cm} p{1cm}}
 \multicolumn{1}{l}{$\log_{10}\zeta$} & \multicolumn{3}{c}{$\log_{10}(\Delta T_{\rm jet}/T_{\rm collapse})$} & \multicolumn{3}{c}{$\log_{10}(U_{\rm jet}/(\Delta p/\rho)^{\frac{1}{2}})$} & \multicolumn{3}{c}{$\log_{10}(\Delta z/R_{0})$} & \multicolumn{3}{c}{$\log_{10}(V_{\rm impact}/V_{\rm max})$} \\ 
 & & & & & & & & & & & & \\
 & c.$\gradp$ & rigid & free & c.$\gradp$ & rigid & free & c.$\gradp$ & rigid & free & c.$\gradp$ & rigid & free \\
 & & & & & & & & & & & & \\
-4.0 & -7.48 & -7.45 & -7.47 & 3.96 & 3.96 & 3.96 & -2.04 & -2.04 & -2.04 & -8.97 & -8.97 & -8.99 \\
-3.9 & -7.30 & -7.30 & -7.30 & 3.86 & 3.87 & 3.86 & -1.97 & -1.98 & -1.97 & -8.77 & -8.77 & -8.79 \\
-3.8 & -7.14 & -7.15 & -7.15 & 3.76 & 3.77 & 3.76 & -1.91 & -1.91 & -1.91 & -8.57 & -8.57 & -8.60 \\
-3.7 & -6.98 & -6.99 & -6.98 & 3.66 & 3.67 & 3.66 & -1.84 & -1.84 & -1.86 & -8.37 & -8.37 & -8.40 \\
-3.6 & -6.82 & -6.83 & -6.81 & 3.56 & 3.56 & 3.56 & -1.78 & -1.78 & -1.78 & -8.17 & -8.17 & -8.20 \\
-3.5 & -6.65 & -6.66 & -6.64 & 3.45 & 3.46 & 3.46 & -1.71 & -1.71 & -1.70 & -7.96 & -7.97 & -8.00 \\
-3.4 & -6.49 & -6.50 & -6.46 & 3.35 & 3.36 & 3.36 & -1.64 & -1.65 & -1.62 & -7.76 & -7.77 & -7.79 \\
-3.3 & -6.32 & -6.33 & -6.29 & 3.25 & 3.26 & 3.26 & -1.58 & -1.58 & -1.55 & -7.56 & -7.57 & -7.59 \\
-3.2 & -6.15 & -6.17 & -6.11 & 3.15 & 3.16 & 3.16 & -1.51 & -1.52 & -1.48 & -7.36 & -7.37 & -7.38 \\
-3.1 & -5.99 & -6.00 & -5.96 & 3.05 & 3.06 & 3.06 & -1.45 & -1.46 & -1.41 & -7.16 & -7.17 & -7.17 \\
-3.0 & -5.82 & -5.84 & -5.80 & 2.95 & 2.96 & 2.95 & -1.39 & -1.39 & -1.35 & -6.96 & -6.96 & -6.97 \\
-2.9 & -5.65 & -5.67 & -5.64 & 2.85 & 2.86 & 2.85 & -1.32 & -1.33 & -1.29 & -6.76 & -6.76 & -6.65 \\
-2.8 & -5.48 & -5.51 & -5.48 & 2.75 & 2.76 & 2.75 & -1.26 & -1.27 & -1.23 & -6.56 & -6.59 & -6.42 \\
-2.7 & -5.32 & -5.34 & -5.32 & 2.65 & 2.66 & 2.65 & -1.19 & -1.20 & -1.17 & -6.36 & -6.36 & -6.19 \\
-2.6 & -5.15 & -5.18 & -5.15 & 2.54 & 2.55 & 2.55 & -1.13 & -1.14 & -1.11 & -6.15 & -6.15 & -5.96 \\
-2.5 & -4.98 & -5.01 & -4.98 & 2.44 & 2.45 & 2.43 & -1.07 & -1.08 & -1.05 & -5.95 & -5.95 & -5.73 \\
-2.4 & -4.81 & -4.83 & -4.80 & 2.34 & 2.34 & 2.34 & -1.01 & -1.01 & -1.00 & -5.75 & -5.75 & -5.49 \\
-2.3 & -4.65 & -4.66 & -4.63 & 2.24 & 2.23 & 2.25 & -0.94 & -0.95 & -0.94 & -5.55 & -5.54 & -5.27 \\
-2.2 & -4.48 & -4.48 & -4.46 & 2.14 & 2.13 & 2.16 & -0.88 & -0.89 & -0.88 & -5.35 & -5.33 & -5.04 \\
-2.1 & -4.31 & -4.31 & -4.29 & 2.04 & 2.02 & 2.07 & -0.82 & -0.83 & -0.83 & -5.15 & -5.13 & -4.83 \\
-2.0 & -4.16 & -4.14 & -4.13 & 1.93 & 1.91 & 1.98 & -0.76 & -0.77 & -0.77 & -4.94 & -4.93 & -4.64 \\
-1.9 & -4.00 & -3.98 & -3.98 & 1.83 & 1.80 & 1.89 & -0.70 & -0.71 & -0.72 & -4.75 & -4.73 & -4.47 \\
-1.8 & -3.85 & -3.82 & -3.82 & 1.73 & 1.69 & 1.80 & -0.64 & -0.64 & -0.67 & -4.56 & -4.54 & -4.34 \\
-1.7 & -3.67 & -3.66 & -3.66 & 1.63 & 1.58 & 1.72 & -0.59 & -0.58 & -0.62 & -4.36 & -4.35 & -4.31 \\
-1.6 & -3.50 & -3.48 & -3.50 & 1.52 & 1.48 & 1.62 & -0.53 & -0.52 & -0.57 & -4.16 & -4.16 & -4.11 \\
-1.5 & -3.32 & -3.24 & -3.32 & 1.41 & 1.38 & 1.53 & -0.47 & -0.46 & -0.52 & -3.93 & -3.83 & -3.77 \\
-1.4 & -3.13 & -2.94 & -3.14 & 1.31 & 1.28 & 1.43 & -0.42 & -0.39 & -0.48 & -3.72 & -3.40 & -3.47 \\
-1.3 & -2.97 & -2.63 & -2.96 & 1.20 & 1.20 & 1.29 & -0.36 & -0.33 & -0.44 & -3.52 & -2.99 & -3.17 \\
-1.2 & -2.81 & -2.32 & -2.79 & 1.10 & 1.12 & 1.15 & -0.31 & -0.26 & -0.40 & -3.31 & -2.62 & -2.89 \\
-1.1 & -2.59 & -2.02 & -2.62 & 1.00 & 1.06 & 1.01 & -0.26 & -0.19 & -0.36 & -2.96 & -2.29 & -2.67 \\
-1.0 & -2.31 & -1.77 & -2.46 & 0.90 & 1.02 & 0.88 & -0.20 & -0.12 & -0.33 & -2.60 & -1.99 & -2.45 \\
-0.9 & -2.01 & -1.58 & -2.32 & 0.81 & 0.99 & 0.77 & -0.15 & -0.06 & -0.30 & -2.23 & -1.72 & -2.32 \\
-0.8 & -1.70 & -1.40 & -2.11 & 0.73 & 0.96 & 0.68 & -0.09 & -0.03 & -0.28 & -1.88 & -1.47 & -2.28 \\
-0.7 & -1.41 & -1.25 & -1.88 & 0.65 & 0.94 & 0.59 & -0.04 & -0.04 & -0.26 & -1.55 & -1.23 & -2.38 \\
-0.6 & -1.14 & -1.10 & -1.55 & 0.58 & 0.92 & 0.49 & 0.02 & -0.07 & -0.22 & -1.25 & -1.06 & -2.14 \\
-0.5 & -0.92 & -1.00 & -1.04 & 0.53 & 0.90 & 0.48 & 0.08 & -0.11 & -0.18 & -0.97 & -0.95 & -1.44 \\
-0.4 & -0.72 & -0.91 & -0.75 & 0.49 & 0.88 & 0.50 & 0.13 & -0.16 & -0.14 & -0.71 & -0.88 & -0.98 \\
-0.3 & -0.58 & -0.85 & -0.69 & 0.44 & 0.85 & 0.55 & 0.19 & -0.21 & -0.11 & -0.48 & -0.83 & -0.76 \\
-0.2 & -0.47 & -0.82 & -0.53 & 0.39 & 0.83 & 0.59 & 0.25 & -0.26 & -0.09 & -0.24 & -0.81 & -0.62 \\
-0.1 & -0.35 & -0.80 & -0.53 & 0.31 & 0.79 & 0.65 & 0.32 & -0.31 & -0.08 & 0.00 & -0.81 & -0.52 \\
0.0 & -0.25 & -0.81 & -0.58 & 0.23 & 0.77 & 0.71 & 0.40 & -0.35 & -0.07 & 0.23 & -0.82 & -0.46 \\
\end{tabular}
\caption{Data from the numerical calculations explained in section~\ref{s:num} and presented in figures~\ref{fig:jettime}-\ref{fig:vtor} for the normalised jet impact time, normalised jet speed, normalised bubble centroid displacement and normalised bubble volume at jet impact as a function of the anisotropy parameter $\zeta$. The data are given for three different jet drivers: constant pressure gradient (c.$\gradp$), rigid surface and free surface.}
\label{tab:tjetujet}
\end{center}
\end{table}

\end{document}